\documentclass[journal=aanmf6,manuscript=letter]{achemso}
\setkeys{acs}{maxauthors=0}

\usepackage[version=3]{mhchem}
\usepackage{mathtools}
\usepackage{amsmath}
\usepackage{amsfonts}
\usepackage{float}
\usepackage{hyperref}
\SectionNumbersOn

\usepackage{chemformula} % Formula subscripts using \ch{}
\usepackage[T1]{fontenc} % Use modern font encodings

\DeclareUnicodeCharacter{2212}{-}

\author{Lin Yang}
\email{linyang@bit.edu.cn}
\affiliation{Department of Physics, McGill University, 3600 Rue University, Montreal, Quebec, H3A 2T8, Canada}
\alsoaffiliation{School of Materials Science and Engineering, Beijing Institute of Technology, Beijing, 100081, China}
\alsoaffiliation{Department of Chemistry, McGill University, 801 Sherbrooke St. West, Montreal, Quebec, H3A 0B8, Canada}
\author{Audrey Moores}
\affiliation{Centre in Green Chemistry and Catalysis, Department of Chemistry, McGill University, 801 Sherbrooke St. West, Montreal, Quebec, H3A 0B8, Canada}
\alsoaffiliation{Department of Materials Engineering, McGill University, 3610 University Street, Montreal, Quebec, H3A 0C5, Canada}
\author{Tomislav Fri\v{s}\v{c}i\'{c}}
\affiliation{Department of Chemistry, McGill University, 801 Sherbrooke St. West, Montreal, Quebec, H3A 0B8, Canada}
\author{Nikolas Provatas}
\email{nikolaos.provatas@mcgill.ca}
\affiliation{Department of Physics, McGill University, 3600 Rue University, Montreal, Quebec, H3A 2T8, Canada}
\alsoaffiliation{McGill High Performance Computing Centre, École de Technologie Supérieure (ETS), 1100 Rue Notre Dame Ouest, Montreal, Quebec, H3C 1K3, Canada}

\title[mechanopfc]{A Thermodynamics Model for Mechanochemical Synthesis of Gold Nanoparticles: Implications for Solvent-free Nanoparticle Production}

\keywords{Mechanosynthesis, Nanoparticles, Phase field method, Structural-phase-field-crystal model, Reaction path}

\begin{document}

\begin{tocentry}

\begin{figure}[H]
	\centering
	\includegraphics[width=1.0\textwidth]{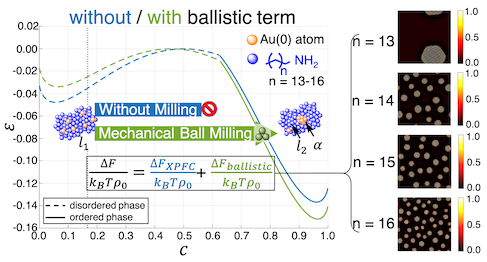}
	\caption{A graphical entry for the Table of Contents}
	\label{FIG:TOC}
\end{figure}

\end{tocentry}

\begin{abstract}
	Mechanochemistry is becoming an established method for the sustainable, solid-phase synthesis of scores of nano-materials and molecules, ranging from active pharmaceutical ingredients to materials for cleantech. Yet we are still lacking a good model to rationalize experimental observations and develop a mechanistic understanding of the factors at play during mechanically assisted, solid-phase nanoparticle synthesis. We propose herein a structural-phase-field-crystal (XPFC) model with a ballistic driving force to describe such a process, with the specific example of the growth of gold nanoparticles in a two component mixture.
	The reaction path is described in the context of free energy landscape of the model, and dynamical simulations are performed based on phenomenological model parameters closely corresponding to the experimental conditions, so as to draw conclusions on nanoparticle growth dynamics. It is shown that the ballistic term lowers the activation energy barrier of reaction, enabling the reaction in a temperature regime compatible with experimental observations. The model also explains the mechanism of precipitated grain size reduction that is consistent with experimental observations. Our simulation results afford novel mechanistic insights into mechanosynthesis with implications for nanaparticle production and beyond.
\end{abstract}

\section{Introduction}

Mechanochemistry focuses on the study of chemical transformations induced and/or sustained by mechanical force in the form of grinding, milling, shearing or other types of mechanical agitation\cite{james2012mechanochemistry, C2CS35442J, doi:10.1021/acs.joc.6b02887, C7GC03797J,C7SC05371A}. It has recently emerged as a versatile route for the synthesis of a wide range of molecules and materials, including organic, pharmaceutical and biomolecular targets \cite{C3GC40302E,C6CC02015A, mottillo2017advances, quaresma2017mechanochemistry}, metal-organic \cite{Friscic2013,C7GC01078H, C8CC03189D, D0CE00091D}, oxides and hybrid materials \cite{B600363J, C7SE00094D, doi:10.1021/acs.accounts.9b00454,C7CS00813A, C2DT30349C, C2CS35462D}, nanoparticle (NP) \cite{ Rak2014, M2017bis}, catalysts \cite{https://doi.org/10.1002/anie.201903545, Ralphs2015, Ralphs2014, Amrute2021} and more\cite{zhu2017solvent, C3CS60052A}.
The ability to avoid the use of bulk solvents and to circumvent solubility-related limitations of conventional synthesis has made mechanochemistry attractive as a tool of sustainable synthesis \cite{C1CS15204A,doi:10.1021/acscentsci.6b00277, fiss2020solvent}. It is also affording access to molecules, materials and reactions that are difficult or even considered impossible to make \cite{gaffet1999some, silva2003structural, MOORES201833, C9GC00304E}.
The ability to use mechanochemistry for the synthesis of NPs has enabled scientific discoveries as well as technological advancement\cite{C3CS35468G}. This concept was first demonstrated  by Geckeler and coworkers, who discovered the controlled growth of AuNPs  (6-28 nm) in the solid phase \cite{B905260G}. A number of examples have since been reported\cite{D0TA05183G, MOORES201833}. In particular, our group has shown the scalable formation of ultra-small AuNPs in the presence of long chain amines as ligands, with excellent NP size control and tunability from the choice of ligand chain length \cite{Rak2014}. This approach was readily applicable to other metal-based and binary systems \cite{fiss2020solvent, C6RA03711A, M2017bis, Sepelak2007, Balaz2003, C3FD00117B}, and the very first real-time mechanistic insights into mechanochemical Au NP formation have been reported by the groups of Camargo and Emmerling through \textit{in situ} X-ray diffraction and absorption spectroscopies \cite{D0CC03862H, D0CE00826E}. The ability to track reactive pathways in situ provides invaluable information regarding mechanochemical methods\cite{Ma2014, Katsenis2015, Batzdorf2015, Kulla2017, Akimbekov2017, Kulla2018, Germann2020}, yet it has limitations in its ability to provide information at timescales relevant for the observed phenomena \cite{Michalchuk2017}.

Despite the growing popularity of mechanochemistry across chemical and materials syntheses, there is very little fundamental understanding on how such transformations take place, and what is the role of mechanical energy input in facilitating them. Recent theoretical efforts have looked at studying the way organic molecules mix in the solid phase \cite{Ferguson2019}. Also organic transformation energetics have been used as a means to track energy input in mechanochemical setups \cite{mckissic2014comparison, Andersen2017}. Yet no such studies were undertaken for NP formation. While models for NP formation in solution, based on homogeneous seeding and nanoparticle growth, have been proposed by LaMer and Dinegar in 1950 \cite{doi:10.1021/ja01167a001} and used extensively during the past two decades \cite{doi:10.1002/anie.201604731}, they are hardly applicable to the solid state reaction environment. This is because the notions of concentrations and supersaturations, which are central to NP growth in dilute solution, can not be translated in the liquid or mixture state, as well as because any model of reactivity by milling needs to account for the role of mechanical energy. Indeed, there is currently no general, thermodynamic model for chemical reactivity under continuous input of mechanical energy. The most popular models of mechanochemical reactivity by milling or grinding focus on microscopic, kinetic effects of mechanical force acting upon hard, abrasive inorganic materials\cite{Colacino2018}. Specifically, the ``hot spot'' model posits short-lived microscopic areas of very high temperature induced by mechanical force\cite{urakaev2000mechanism, chen2010measuring}, while in the ``magma-plasma'' model the role of mechanical force is to produce local dislocations and plastic deformation, as well as bond rupture and crystallographic defects\cite{boldyrev1971mechanochemistry, gutman1998mechanochemistry, james2012mechanochemistry}. Melt-driven mechanisms have been proposed\cite{Humphry-Baker2016}. While these models can provide a qualitative view of the effects of mechanical force on bulk solids \cite{butyagin1971kinetics}, they do not provide a route to incorporate the effects of mechanical force into the overall thermodynamic framework of chemical reactivity by milling or grinding. Moreover, the described models have been developed in the context of hard inorganic solids, and are not likely to be applicable to softer systems such as organic, metal-organic or NP synthesis. On one hand, this is evident in the unexpectedly high temperature dependence of mechanochemical reaction kinetics, reported by different groups. On the other, the high temperatures anticipated by the hot spot theory are expected to degrade organic ligands or even cause gold melting, which is incompatible with experimental evidence. Consequently, researchers proposed novel models to account for kinetics in mechanochemical reactions \cite{D0CP01658F}, yet there are currently no models that describe the spatio-temporal properties of nucleation and nanocrystalline precipitate evolution in mechanochemical, solid state, reactions.

The {\it phase field crystal (PFC)} \cite{PhysRevB.75.064107} method is a field theoretic approach whose form is motivated from classical density functional theory (CDFT) \cite{PhysRevB.19.2775}, but which typically replaces microscopic parameters arising from the original CDFT with effective parameters that make spatio-temporal microstructure simulations tractable on diffusional times scales. The starting point of a PFC model is a free energy of a system expressed in terms of a local order parameter field. The free energy captures the local thermodynamic driving forces that characterize reactions between possible phases in the system, as well the energy cost associated with nanoscale interfaces between phases.  Microstructure dynamics is described by the evolution of the order parameter field, driven by the dissipative dynamical minimization of said free energy, subject to mass conservation. \citeauthor{Greenwood2011} extend the original PFC method to the so-called ``structural'' phase field crystal (XPFC) \cite{Greenwood2011} method, where the ``X'' emphasizes the use of specific density-density correlation functions in the free energy to control the crystal symmetries that emerge during crystallization. PFC and XPFC methods has been successfully used in recent years in the description of solidification \cite{PhysRevLett.103.035702, van2009derivation}, structural phase transitions in pure and alloy materials \cite{PhysRevLett.105.045702, PhysRevE.83.031601}, clustering and precipitation \cite{fallah2012phase, Smith2017}.  These phase field crystal models are based purely on thermal energy input to drive reactions. Recently, \citeauthor{Ofori-Opoku2012} \cite{Ofori-Opoku2012} incorporated into an XPFC model a procedure to treat external neutron bombardment by including a \textit{ballistic energy term} developed by \citeauthor{Enrique2000} \cite{A.Enrique1999, Enrique2000} and used it to study microstructural stability in irradiation-driven nanocrystalline systems \cite{Ofori-Opoku2012}.

Here, we proposed a ``thermodynamics-based field theoretic model'' inspired by an XPFC framework (Ref.~\citenum{Ofori-Opoku2012}) to gain mechanistic insight into mechanochemical NP synthesis. This model, for the first time, is used to describe the spatio-temporal evolution of microstructure during ball milling mechanochemistry, and provides general understandings into how ballistic mechanical energy affects the milling or grinding process. In doing so, we considered two necessary modifications arising from the specific nature of NP mechanosynthesis: the form of driving force (neutron irradiation vs mechanical input), and the reaction environment. In particular, the ballistic term was altered to mimic the mechanical milling taking place while growing \ce{Au}NPs by reduction of \ce{HAuCl4} in a stainless steel ball milling apparatus\cite{Rak2014}. Also, the XPFC model was adapted to account for the effect of ligand type (mass and length) as the NP grow in the presence of excess long-chain amine ligands.
The model was used to elucidate the mechanism of \ce{Au}NPs precipitation and growth in the solid phase. This was made possible by the newly developed XPFC ballistic term, which provided for the first time a framework for tuning the milling frequency and the ligand type. An analysis of the model's thermodynamics helped elucidate how mechanical energy delivery affects the free energy of the reaction, and thus how a NP precipitation reaction is feasible under normal room-temperature experimental conditions. Moreover, spatio-temporal microstructure evolution using the XPFC model made it possible to simulate the nucleation and growth of nanoparticles in the solid phase and to analyze their size. Our results reproduced experimental trends, offering a deeper understanding of potential mechanisms for solid-state NP growth.

\section{XPFC Model for Nanoparticle Mechanosynthesis}
\label{sec:model}

\begin{figure}[H]
	\centering
	\includegraphics[width=1.0\textwidth]{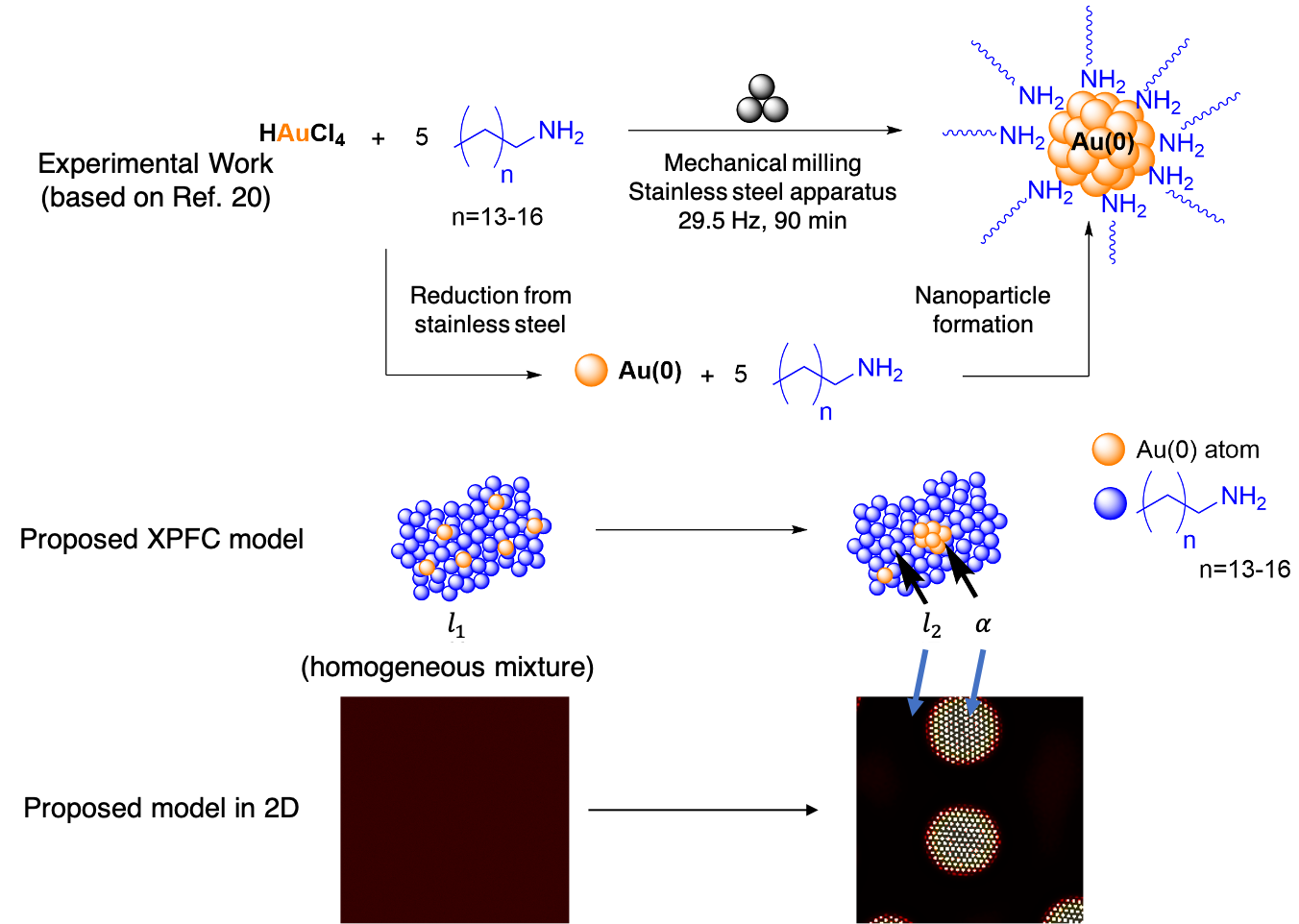}
	\caption{Schematic representation of the mechanochemical synthesis of AuNPs based on experimental work \citenum{Rak2014}, the proposed XPFC model, and simulations of the model in 2D.}
	\label{FIG:1}
\end{figure}

In our previous work, we have shown that AuNPs could be obtained with excellent size control by reacting with an \ce{Au} precursor, hydrochloroauric acid, with long chain amines, in the solid phase, by milling them in a stainless steel reactor\cite{Rak2014}. This reaction can be viewed as a two step process with first the reduction of Au(III) species into atomic Au(0), and second the assembly of atomic Au(0) into nanoparticles (Fig. \ref{FIG:1}). We elected to model the second step, namely the formation of AuNPs as a precipitation phase transition, described in a two component (Au(0)-ligands) monotectic system. This modelled reaction is denoted $l_{1} \rightarrow \alpha + l_{2}$, where $l_1$ (e.g. Au(0)-ligands mixture before reaction) and $l_2$ (e.g. Au(0)-ligands mixture after reaction) are two disordered phases, and $\alpha$ (e.g. AuNPs) is the precipitated solid phase. This describes solid precipitation from disordered phase $l_{1}$ to a solute-poor crystalline phase $\alpha$ and a solute-rich disordered phase $l_{2}$. In the developed XPFC model, we used $c$ as the local composition at a given point in space, where $c=0$ at locations solely composed of ligands, and $c=1$ in regions of pure Au phase. The parameters of our model free energy are chosen such that the crystalline phase of Au forms at composition $c=1$. As purely atomistic molecular treatment could not be handled in the XPFC formalism, the reduction of Au(III) to Au(0) was not considered, so as to focus on the role of Au(0) in Au NP formation.
The model is studied in 2D in order to enable faster computer simulation efficiency but without losing the model's ability in elucidating the salient mechanisms of Au NP formation. We chose  the <111> direction to represent NP crystallography, leading to hexagonal closed packed atomic packing of Au crystals in 2D.
%%%%%%%%%%%%%%%%%%%
%%%%%%%%%%%%%%%%%%

The energy functional for the XPFC model for nanoparticle mechanosynthesis (XPFCNanoMecha) was established using the monotectic mixture XPFC model developed by \citeauthor{Smith2017} \cite{Smith2017} as a starting point. The latter model though only considers thermal energy input, so we proceeded with adding another term called ``ballistic term'', following the approach of  \citeauthor{Enrique2000} \cite{A.Enrique1999,Enrique2000}. The resulting energy functional is provided here:
\begin{align}
	\frac{\Delta\mathcal{F}\left[n(\vec{x}),c(\vec{x})\right]}{k_{B}T\rho_{0}} & = \left(\frac{\Delta\mathcal{F}_{id}\left[n(\vec{x})\right]}{k_{B}T\rho_{0}} + \frac{\Delta\mathcal{F}_{mix}\left[n(\vec{x}),c(\vec{x})\right]}{k_{B}T\rho_{0}} + \frac{\Delta\mathcal{F}_{ex}\left[n(\vec{x}),c(\vec{x})\right]}{k_{B}T\rho_{0}}\right) + \frac{\Delta\mathcal{F}_{ballistic}\left[n(\vec{x}),c(\vec{x})\right]}{k_{B}T\rho_{0}} \nonumber \\
	                                                                           & = \frac{\Delta\mathcal{F}_{XPFC}\left[n(\vec{x}), c(\vec{x})\right]}{k_BT\rho_0} + \frac{\Delta\mathcal{F}_{ballistic}\left[n(\vec{x}),c(\vec{x})\right]}{k_{B}T\rho_{0}}
	\label{eq:freefunctional}
\end{align}
where $k_{B}$ is Boltzmann's constant, $T$ is the temperature, and $\rho_{0}$ is the total system reference density. The free energy functional has been broken into the ideal $\Delta\mathcal{F}_{id}\left[n(\vec{x})\right]$, mixing $\Delta\mathcal{F}_{mix}\left[n(\vec{x}),c(\vec{x})\right]$, excess $\Delta\mathcal{F}_{ex}\left[n(\vec{x}),c(\vec{x})\right]$ and ballistic $\Delta\mathcal{F}_{ballistic}\left[n(\vec{x}),c(\vec{x})\right]$ parts, respectively. The function $n(\vec{x})$ is a normalized term for the density, defined as $n(\vec{x})=\left(\rho(\vec{x})-\rho_0\right)/\rho_0$ where $\vec{x}$ denotes the position, $\rho(\vec{x})$ is the local mass density, and $\rho_0$ is the reference mass density. The function $c(\vec{x})$ is the local composition, as defined above, which varies with respect to position $\vec{x}$. By adopting a series of reasonable approximations (which are described in the supplementary information) for each term, the free energy form is developed in terms of $\{n(\vec{x}), c(\vec{x})\}$. To simplify mathematical expressions, a dimensionless free energy per volume $V$ is often introduced in PFC models, given by
\begin{align}
	\mathcal{E} =\frac{\Delta\mathcal{F}}{k_{B}T\rho_{0}V}
\end{align}

\subsection{Incorporating ballistic energy in the XPFC model}

The ballistic term $\Delta\mathcal{F}_{ballistic}$ accounts for the energy dissipated into the system by the external mechanical force applied during a mechanochemical reaction, which is expressed as $\Delta\mathcal{F}_{ballistic}=\gamma \mathcal{G}$, where $\mathcal{G}$ accounts for mechanically induced interactions between parts of the system, and $\gamma = {\Gamma}/{M}$ where $\Gamma$ is the scaled forcing frequency on system particles while $M$ is the particle mobility. It is worth mentioning that $\gamma$ is a phenomenological parameter in this model, as microscopic information on the energy input associated with each ball hitting event is not easy available. Adapting the work of \citeauthor{Enrique2000},\cite{Enrique2000} the specific form of $\mathcal{G}$ proposed in this work is
%%%%%%%%%%%%%%%%

\begin{align}
	\mathcal{G} = \frac{1}{2}\int d\vec{x} \, d\vec{x}'\, \Pi\left(c(\vec{x})\right)\, n(\vec{x}) \, g(\vec{x}-\vec{x}') \, n(\vec{x}')
	\label{eq:G}
\end{align}
%%%%%%%%%%%%%%%%%
where $\Pi(c(\vec{x}))$ modulates the mechancial energy dissipated into the system at a point $\vec{x}$ (represented by the integral in Eq.~\ref{eq:G}) as a function of the local composition $c(\vec{x})$. The kernel for the ballistic exchange $g$ satisfies a Poisson's equation, i.e.,
%%%%%%%%%%
\begin{align}
	\nabla^2 g(\vec{x}-\vec{x}') = -\left[\delta(\vec{x}-\vec{x}')-w_{R}(\vec{x}-\vec{x}^{\prime})\right],
	\label{eq:kernel}
\end{align}
%%%%%%%%%%%%%%%%
where the source delta function $\delta(\vec{x}-\vec{x}')$ accounts for short-range ballistic interactions between particles, while the weight function $w_{R}(\vec{x}-\vec{x}')$ accounts for interactions between points in the system dissipated over longer spatio-temporal scales, the form of which depends on the energy dissipation length $R$. The purpose of this construction strategy for $\Delta\mathcal{F}_{ballistic}$ term can be more clearly depicted in the phase field dynamics, which obey the conserved Langevin equations:
\begin{align}
	\frac{\partial n}{\partial t}  & = M\nabla^2\frac{\delta \Delta\mathcal{F}_{XPFC}}{\delta n} - \Gamma\left(n-\langle n \rangle\right) \\
	\frac{\partial c}{\partial t}  & = M\nabla^2\frac{\delta \Delta\mathcal{F}_{XPFC}}{\delta c} - \Gamma\left(c-\langle c \rangle\right)
\end{align}
where $\langle n \rangle$ and $\langle c \rangle$ are the \textit{ideal} average density and composition field values under ballistic energy, respectively. Once $n$ or $c$ deviates from the average value, external driving force would be applied. The $\langle n \rangle$ and $\langle c \rangle$ forms correspond to spatial averages of the density and composition fields:
\begin{align}
	\langle n \rangle & = \int d\vec{x}' w_{R}(\vec{x}-\vec{x}^{\prime}) n(\vec{x}') \\
	\langle c \rangle & = \int d\vec{x}' w_{R}(\vec{x}-\vec{x}^{\prime}) c(\vec{x}')
\end{align}
Further details on $w_{R}(x:=|\vec{x}-\vec{x}^{\prime}|)$ are available in supplementary information.

\subsection{The interpolation function $\Pi(c)$}

The interpolation function $\Pi(c)$ allows us to modulate the amount of ballistic energy input into the system as a function of the local distribution of particles, which in turn is represented in the PFC model by the local composition $c(\vec{x})$.  Specifically, the energy absorbed by a ligand-rich volume element is different from that absorbed by a gold-rich volume. This can be quantified by considering the an elastic collision analogy wherein  a stationary ball with a small mass $m$ is hit by a heavy ball of $M$ (${m}/{M}\ll 1$) moving at speed $\vec{u}$. After the collision, the small ball will move at a velocity of $\vec{v} = \left[{2m}/{(M+m)}\right]\vec{u}$ and a kinetic energy
%%%%%%%%%%%%%%%%%%
\begin{align}
	E_{m} =\frac{2 m}{1 + m/M}\,|\vec{u}|^2 \approx 2 m |\vec{u}|^2
\end{align}
%%%%%%%%%%%%%%%%%%%
In mechanosynthesis, the milling ball (ball milling process) or the pestle (grinding process) can be regarded as the ``heavy ball'', while the particles can be regarded as instances of the ``light ball''. The Au and ligand particles can be regarded as the ``light balls'' in the present model for our targeted experiment, denoting the unit mass of Au as $m_{gold}$, and that of ligand as $m_{ligand}$, thus, it is reasonable to approximate the energy input locally into a small volume of the system as being proportional to the integrated mass around the small local volume, modulated by the type of mass via the composition $c$ (the composition of gold) according to
%%%%%%%%%%%%%%%%
\begin{align}
	\Pi(c) = 1 - \left(1 - \frac{m_{gold}}{m_{ligand}}\right)c
	\label{Pi_form}
\end{align}
%%%%%%%%%%%%%%%%%%
In Eq.~(\ref{Pi_form}), $c \rightarrow 0$ implies that a local volume centred at $\vec{x}$ is occupied mostly by ligand molecules, and the total mechanical energy represented by the integral in Eq.~(\ref{eq:G}) is transferred to ligand molecules at position $\vec{x}$. On the other hand, the limit $c \rightarrow 1$ implies that the local volume is occupied mostly by gold atoms, and the fraction ${m_{gold}}/{m_{ligand}}$ of the total mechanical energy is transferred to gold atoms at $\vec{x}$.  This behaviour of $\Pi(c)$ guarantees that the energy gained is proportional to both the mass of the local area and the type of atomic species.
%%%%%%%%%%

\subsection{Full form of XPFC model}

Combining all the mechanisms discussed above, the complete free energy functional of the structural-phase-field-crystal (XPFC) model describing reactions in a monotectic alloy with ballistic energy input is given by
%%%%%%%%%%%%%%
\begin{align}
	\frac{\Delta\mathcal{F}[n,c]}{k_{B}T\rho_{0}} = \int d\vec{x} & \left\{ \frac{n(\vec{x})^2}{2}-\eta\frac{n(\vec{x})^3}{6} + \chi\frac{n(\vec{x})^4}{12} +  \frac{\omega}{2}\epsilon(T)\cdot (c-c_0)^2+
	\frac{W_c}{2}|\vec{\nabla}c|^2\right. \nonumber                                                                                                                                                                             \\
	                                                              & - \frac{1}{2}n(\vec{x})\int d\vec{x}' \left[\,C_{eff}\left(\vec{x}-\vec{x}'\right)-\gamma\Pi(c)g\left(\vec{x}-\vec{x}'\right)\,\right]n(\vec{x}') \nonumber \\
	                                                              & + \left. \omega\left(n(\vec{x})+1\right)\cdot\left[c\,\ln\left(\frac{c}{c_0}\right)+(1-c)\,\ln\left(\frac{1-c}{1-c_0}\right)\right] \right\}
	\label{eq:freeballistic}
\end{align}
where $\eta, \chi, \omega$ are constants that set the scale of the free energy in the bulk phases, $W_c$ sets the scale of compositional fluctuations, $\epsilon(T)$ controls the enthalpy of mixing as a function of temperature ($T$), $C_{eff}(\vec{x}-\vec{x}^\prime)$ controls the emergence of crystal ordering from a disordered phase and $c_o$ is a reference concentration. Further details on these parameters is found in the supplemental information.

\section{Model Outputs}

\subsection{Kinetics and thermodynamics outputs}
\label{sec:equilibrium_props}

From the expression of the free energy function in Eq.~(\ref{eq:freeballistic}), we could extract expressions for the bulk free energy on the various phases, from which we could draw conclusions about the kinetics of NP precipitation. The derivation of the free energies of the bulk phases follows a simple type of coarse graining procedure developed by us in previous work \cite{Provatas2010} and described in the supplementary information. The explicit form of the bulk free energy of the disordered phase (denoted $\mathcal{E}_{d}$) and the ordered hexagonal phase (denoted $\mathcal{E}_{o}$) become, respectively,
%%%%%%%%%%%%%%%%%%%%%%%%%%
\begin{align}
	\mathcal{E}_{d} & = \left(1-\hat{C}_{eff}(0)+ \gamma\hat{g}(0)\Pi(c)\right)\frac{n_0^2}{2} - \eta\frac{n_0^3}{6} + \chi\frac{n_0^4}{12} + \omega\Delta f_{mix}(n_0, c) \label{eq:edisorder}            \\
	\mathcal{E}_{o} & = \mathcal{E}_d + 3\left(1-\eta n_0 + \chi n_0^2 - \hat{C}_{eff}(k_a) + \gamma\hat{g}(k_a)\Pi(c)\right)\phi_0^2 - 2\left(\eta - 2\chi n_0\right)\phi_0^3 + \frac{15}{2}\chi\phi_0^4,
	\label{eq:eorder}
\end{align}
%%%%%%%%%%%%%%%%%%%%%%%%%%%
where $n_o$ is the average density of a bulk phase and $c$ its average concentration, $\phi_0$ is the mode amplitude which minimizes the free energy by solving $\frac{\partial\mathcal{E}}{\partial\phi}=0$, (where $\phi$ is the mode amplitude described in S.I. Eq. (14)), while $\Delta f_{mix}$ is a shorthand notation for the expression
%%%%%%%%%%%%
\begin{align}
	\Delta f_{mix} \coloneqq \left[n(\vec{x})+1\right]\cdot\left[c\,\ln\left(\frac{c}{c_{0}}\right) + \left(1-c\right)\,\ln\left(\frac{1-c}{1-c_{0}}\right)\right] + \frac{1}{2}\epsilon(c-c_{0})^2
\end{align}
%%%%%%%%%%%%%%%%%%%%%
and $\hat{C}_{eff}(k_a)$ is the Fourier transform of the effective density-density correlation function evaluated at the magnitude of reciprocal lattice vector of the lattice peak, given by $k_a=\frac{2\pi}{1.0}$. Also, $\hat{g}(k)$ is the Fourier transform of external energy interaction kernel defined above.

Phase equilibrium between  $\mathcal{E}_o$ (gold) and  $\mathcal{E}_d$ (gold-ligand mixture) is found from Eq.~\ref{eq:edisorder} and Eq.~\ref{eq:eorder} following well known minimization procedures.\cite{PhysRevB.75.064107, PhysRevLett.88.245701} Namely, the free energy $\mathcal{E}_o$ is minimized with respect to $\phi_0$, and the solution is substituted back into Eq.~\ref{eq:eorder}, yielding a free energy $\mathcal{E}_o$ that depends on the average density $n_0$ and composition $c$. Then, both $\mathcal{E}_d$ and $\mathcal{E}_o$ are compared in these variables using a {\it common plane construction}. In this work, we assumed that the average density of bulk $\mathcal{E}_o$ and $\mathcal{E}_d$ are equal, thus reducing the common plane construction to the usual {\it common tangent construction} in the variable $c$.

Through their parameters, $\mathcal{E}_o$ and $\mathcal{E}_d$  (and hence the equilibrium phase properties) depend on temperature $T$, the mass ratio between the two components, the reduced forcing frequency $\gamma$, and force dissipation length $R$, all of which must be approximated. Here, we adopt the same parameters for the base XPFC free energy as the monotectic system studied by \citeauthor{Smith2017} in Ref. \citenum{Smith2017}. Specifically, we set $\eta=2.0$, $\chi=1.0$, $\omega=0.3$, $\epsilon_0=30$, $T_c=0.15$, and $c_0=0.5$. The temperature is set to  $T=0.06$.
This treatment of the model thus allows us to draw the energy landscape of the system and calculate kinetic and thermodynamic parameters for the reaction $l_{1} \rightarrow \alpha + l_{2}$.

\subsection{Microstructural outputs}
\label{sec:dynamics}

It addition to thermodynamic analysis, the XPFC model also allowed us to track microstructure evolution during mechanochemical synthesis. In particular, we  followed the dynamical aspects of the precipitation of AuNPs from a ligand-gold monotectic system, driven by external mechanical energy (reaction $l_{1} \rightarrow \alpha + l_{2}$).

The dynamical evolution of the $n(\vec{x},t)$ and $c(\vec{x},t)$ fields in the structural-phase-field-crystal model is assumed to obey conserved Langevin equations for each field, which is described in detail in supplementary information.

\section{Implications of the Model}

The  model developed for this work was used to understand two key aspects of the mechanosynthesis of AuNPs with long chain amines\cite{Rak2014}. The first was to test the accuracy of the ballistic term in accounting for the role of mechanic force in this reaction. The second goal was to investigate and rationalize the model's implications on how the amine chain length controls the size of the final AuNPs in the reaction.

\subsection{Role of ballistic term}
\label{sec:onoff}

\subsubsection{How the ballistic term helps nanoparticle precipitation kinetically and thermodynamically}

\begin{figure}[H]
	\centering
	\includegraphics[width=1.0\textwidth]{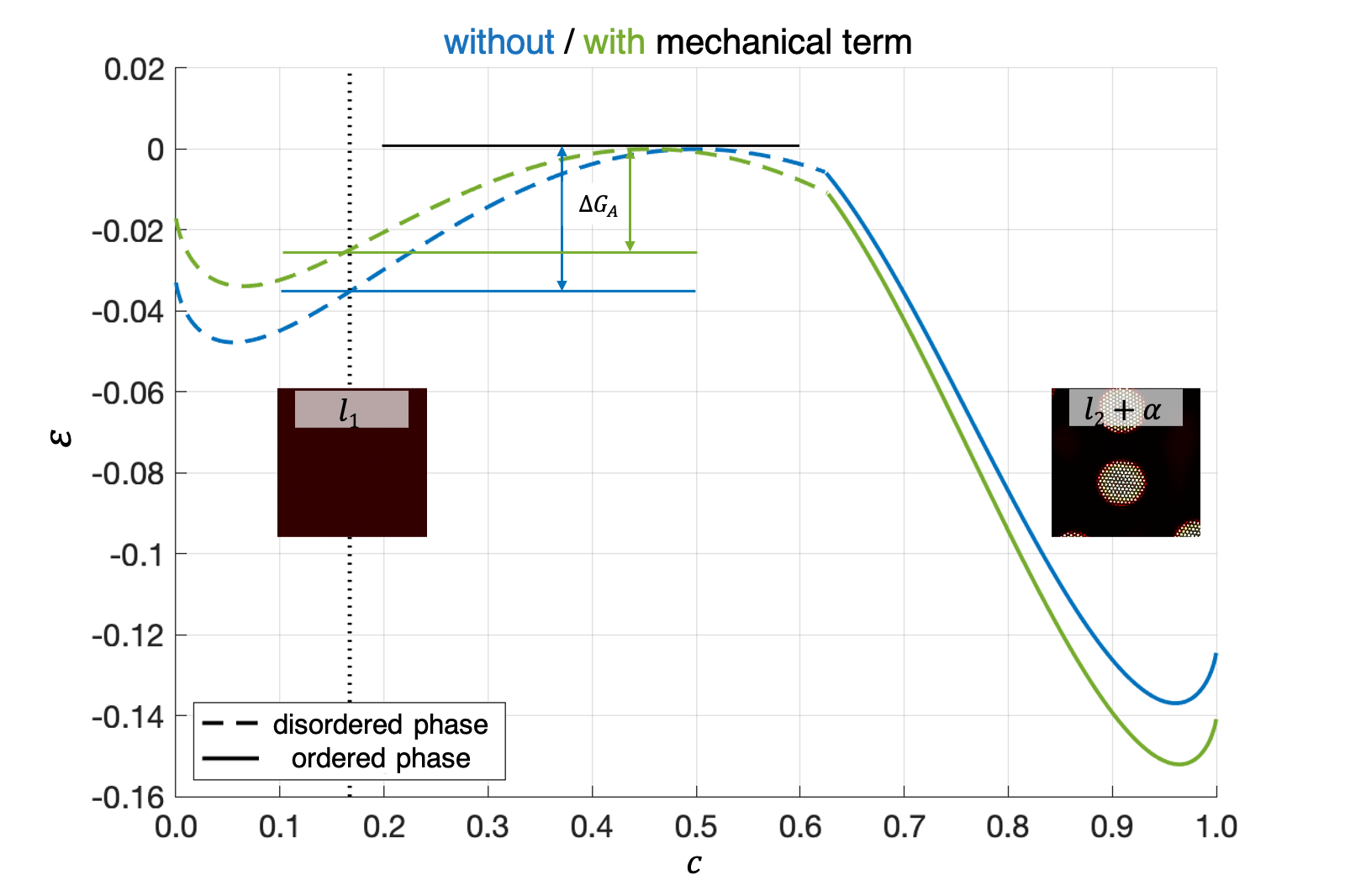}
	\caption{(Colour online) Dimensionless free energy landscape in $\mathcal{E} \times c \times \gamma$ space of a Au(0) - octadecylamine (C18) system. The blue curves correspond to $\gamma = 0$ and the green curves correspond to $\gamma = 0.1$.  The free energy landscapes for both $\gamma$ have been subtracted by their maximum value, aligning them all at $\mathcal{E}=0$. The black vertical dash line at $c=0.167$ marks the reaction starting point. The difference in the green and blue vertical lines indicates $\Delta G_a$, the  energy change required (relative to the maximum energy) to activate the precipitation of the ordered phase from the disordered phase.}
	\label{FIG:2}
\end{figure}

The role of externally induced mechanical forces on the free energy of the system was studied through Eq.~\ref{eq:edisorder} and Eq.~\ref{eq:eorder}. When there is no induced mechanical force $\gamma=0$, the free energy of each phase reduces to that described by usual thermal equilibrium conditions. When there is induced mechanical force ($\gamma > 0$), free energy landscape of each phase is altered. Importantly, both the $\gamma\hat{g}(0)\Pi(c)$ term and the temperature dependent correlation term $\hat{C}_{eff}(0)$ are modulated by the square of the average density $n_0^2$ in the disordered phase (Eq.~\ref{eq:edisorder}), while $\gamma\hat{g}(k_a)\Pi(c)$ and $\hat{C}_{eff}(k_a)$ are coupled to the square of the order parameter amplitude $\phi_0^2$ in the ordered phase (Eq.~\ref{eq:eorder}). Thus, with $\gamma>0$, the symmetry breaking term $\gamma\hat{g}(0)\Pi(c)$ increases the free energy of the disordered phase (initial stage) relative to its value with $\gamma=0$ in proportion to $\Pi(c)$. Meanwhile, in the ordered phase (formed AuNPs), the term $\gamma\hat{g}(k_a)\Pi(c)$ {\it decreases} the free energy relative to its $\gamma=0$ value in proportion to $\Pi(c)$. Fig.~(\ref{FIG:2}) shows the free energy profile of the system as a function of $c$ and thus as it transitions from initial disordered stage to final AuNPs stage, with $\gamma=0$ and $\gamma=0.1$.

%%%%%%%%%%%%%%%%
%%%%%%%%%%%%%%

Fig.~(\ref{FIG:2}) confirms visually the trends explained above. Specifically, the reaction is favoured, both kinetically and thermodynamically by the ballistic energy input ($\gamma>0$). The time to nucleation $t_n$ can be evaluated by estimating the activation barrier $\Delta G_a$ according to
\begin{align}
	t_n = t_re^{-\frac{\Delta G_a}{k_BT}}
\end{align}
where $t_r$ is a reference time which estimates the time that an atom statistically requires to hop to from the disordered phase ($l_2$) to the solid phase ($\alpha$).
% {\color{red} disordered phase ($l_2$) to the solid phase ($\alpha$).}

Without the ballistic term, the activation barrier is $\Delta G_a = 0.0354$, while with $\gamma=0.1$, the activation barrier reduces to $\Delta G_a = 0.0250$, which represents a percentage reduction of $f=29.28\%$ of the nucleation barrier compared to its value in the absence of ballistic energy. Denoting the time to nucleation with the ballistic term as $t_{b}$ and that without the ballistic term as $t_{0}$, $t_{0}=\left(\frac{t_b}{t_r}\right)^ft_{b}$. Considering a typical reference time $t_r=1$ ns and a typical mechanosynthesis reaction of $t_b=90$ min (as described in Ref. \citenum{Rak2014}) predicts an acceleration of the reaction rate by a factor of 5346, which would take 334 days for the reaction to occur without mechanical activation. It is noteworthy that the value of $\gamma$ was chosen arbitrarily here as it is a phenomenological parameter. However, the larger $\gamma$, the more $\Delta G_a$ is reduced. This analysis thus suggests that a plausible mechanism for nanoparticle mechanosynthesis is that ballistic energy reduces the barrier to precipitation. It is also importantly to note that while these shifts in relative stability of the ordered phase relative to the disordered phase mimics an effective ``temperature quench'', the induced mechanical force \emph{cannot} be regarded as a higher ``effective'' temperature.

\subsubsection{Impact of the ballistic term on nanoparticle nucleation and growth}

The precipitation process was simulated with the ballistic energy term off or on, to simulate the reaction without or with the mechanical force, respectively.
This group of dynamical simulations consists of three samples, where the ligand was chosen to be heptadecylamine (C17). The \ce{Au(0)} : ligand ratios are all set to be $1:5$ as in the reference experimental work\cite{Rak2014}. The initial state of the system in each case was modelled by uniform spatial distributions for $n$ and $c$, which emulates a randomly fluctuating disordered starting phase of the system. In the Sample~1 simulation, the ballistic term was deactivated using $\gamma=0$ and $R=0$, and sets the noise according to S.I. Eq. (17) to simulate the reaction without mechanical force. The Sample~2 simulation is the same in every way to that in Sample~1, except the ballistic term is activated using $\gamma = 0.1$ and $R = 9.7a$. A third simulation was also performed (called Sample~3 simulation), in which the ballistic term was initially the same as that of Sample~2 simulation but was deactivated after $430$ numerical timesteps. This was done to simulate the role of partially injecting external energy for a brief time, after which it is turned off during the remainder of an ageing process. Ageing has not been reported for AuNPs, but has been shown to be an important step in the mechanosynthesis of \ce{Bi2S3}~\cite{M2017bis} and \ce{Ni2P}NPs~\cite{fiss2020solvent}. All microstructure simulations were performed with a dimensional resolution of $\Delta x = 0.125$, and the numerical timesteps were set to $\Delta t=0.00015$. Simulations were performed using a semi-implicit Fourier method described in Ref.~\citenum{Provatas2010} on a uniform two-dimensional numerical mesh of $1000 \times 1000$ two-dimensional grid points. The average system density was set to $n_0=0.05$, while the mobility of the density and composition fields were set to $M_n = 1.0$, $M_c = 1.0$ for simplicity. The temperature was fixed at $T=0.06$.

Importantly that the variables we used are dimensionless. Length resolution $\Delta x$ is scaled with respect to the lattice constant of gold, time resolution $\Delta t$ with respect to the inverse atomic mobility of atoms hopping between matrix and nuclei, and energies with respect to $k_B T \rho_0$. The system average density and inverse mobility are presently not known experimentally, with the  mobility being the quantity that is expected to be most challenging to measure or somehow ascertain. Similarly, there is no information to our knowledge about the collision energy transfer coefficient $\Gamma$ . As a result of these considerations, the parameters scaling of our model physical components are necessarily  phenomenological, and further research will be required to calibrate them and compare the output measurements of our model to experiments quantitatively. The purpose of this first work was to propose and validate a plausible physical \textit{mechanisms} at play during nanoparticle precipitation in basic ball milling processes.  We thus focused our attention to examining the role that changing milling energy has on the activation barrier for precipitation and on the particle growth scaling. We then compared them to experiments qualitatively even if a precise quantitative comparison is not yet possible. It is also pointed out that the highly non-linear nature of the model precludes a simple regression-type analysis to fit its coefficients on to given quantities (assuming these were even known experimentally as discussed above). Work is presently under way to use machine learning to tune parameters of such models, but this is still in the early stages.

\begin{figure}[H]
	\centering
	\includegraphics[width=1.0\textwidth]{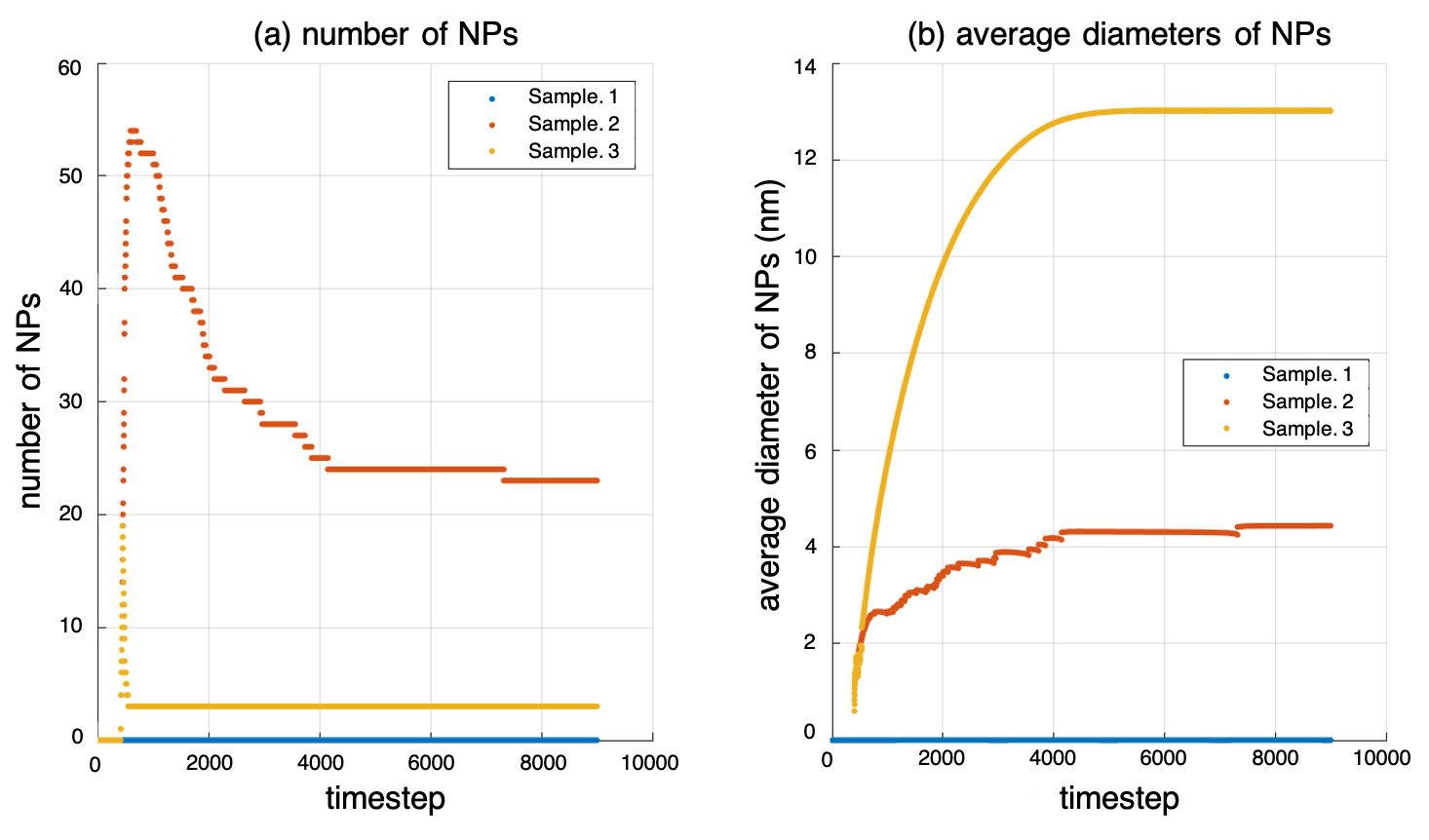}
	\caption{(Colour online). Data comparing numbers of precipitated particles (left) and their average sizes (right) for three different cases of ballistic driving (see text for details).}
	\label{FIG:3}
\end{figure}

Fig.~\ref{FIG:3} compares data from the three aforementioned simulation samples of microstructure evolution. Comparing Sample~1 with Sample~2, it is seen that no AuNPs are precipitated for a considerable period of time (Sample~1), in contrast to the rapid precipitation that occurs when the ballistic term is activated (Sample~2). An interpretation for this is that even in a well-mixed sate, the precipitation will not occur because the activation energy barrier is too high in this case. As shown in the analysis of Fig.~\ref{FIG:2}, turning on the ballistic energy  serves to lower the activation energy barrier, which facilitates the precipitation of \ce{Au} nano-particles through normal thermal fluctuations. Comparison of Sample~2 with Sample~3 was found to give different morphology for the final precipitated products. Since in Sample~3 simulations the ballistic term is deactivated at timestep $=430$, no new nuclei can form after that (and some nuclei also dissolve into the disordered matrix), leaving the nuclei remaining to continue to grow, thus attaining larger NPs sizes compared to NPs in Sample~2 where new nuclei continue to form in the matrix past $=430$ timesteps.
%%%%%%%%%%%%%%%%%%%
%%%%%%%%%%%%%%%%%%

\begin{figure}[H]
	\centering
	\includegraphics[width=1.0\textwidth]{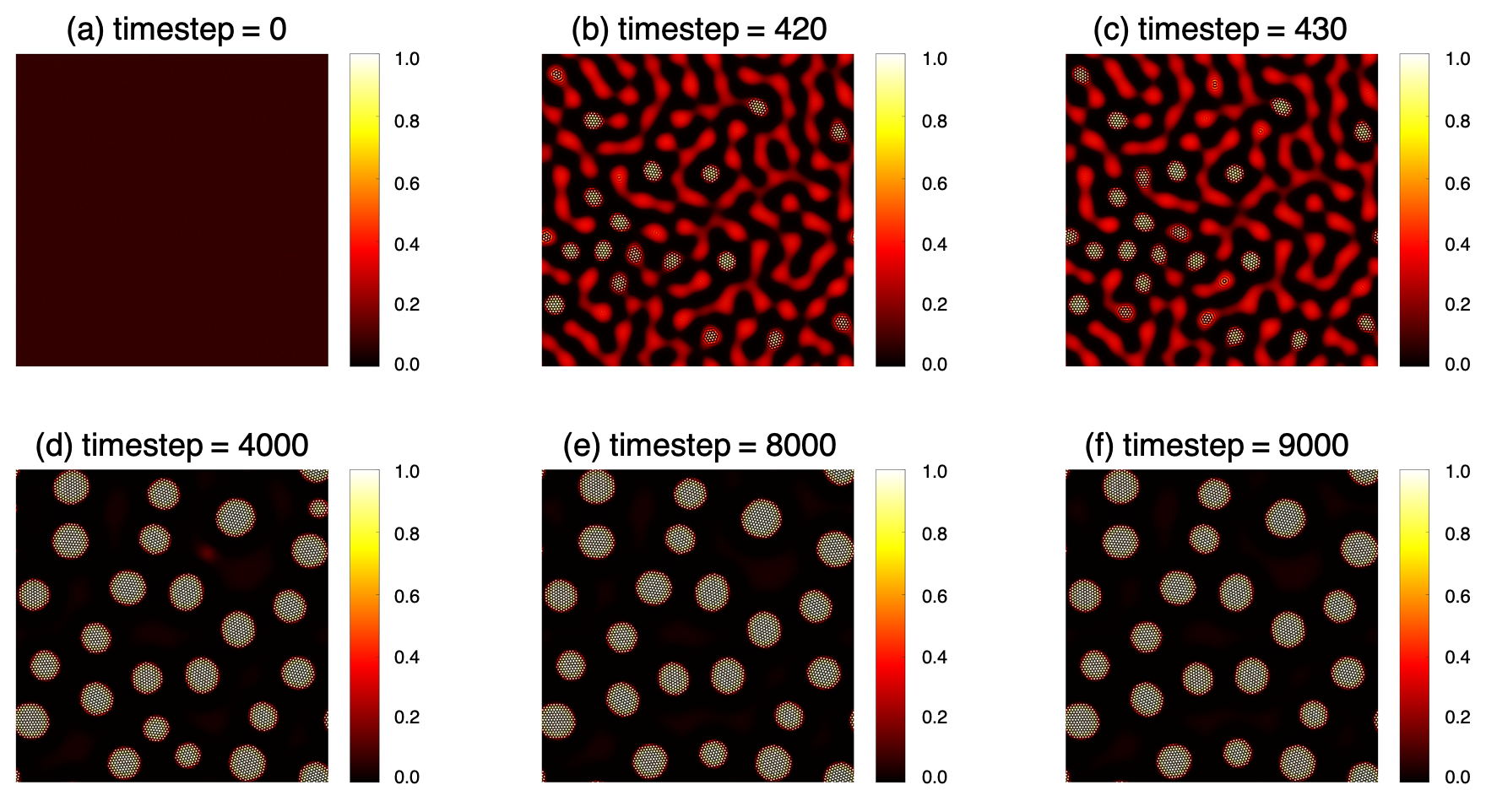}
	\caption{(Colour online) Simulated mechanosynthesis of \ce{Au}NPs from a disordered heptadecylamine (C17) -- \ce{Au(0)} matrix, with the ballistic term continuously active ($\gamma = 0.1$ and $R = 9.7a$). The black / white pixels represent the dimensionless density $n$ and the colourbar readings represent the composition of Au. (a) The initial ${\rm timestep}=0$ configuration is randomly distributed with average reduced density $n_{timestep=0}=0.05$ and composition $c_{timestep=0}=0.167$. (b-c) show the nucleation  process. (d-f) show the post-nucleation growth of NP precipitates.}
	\label{FIG:4}
\end{figure}

Fig.~\ref{FIG:4} shows the result of the microstructure simulations in the Sample~2 case. In the early stages of NP synthesis (figures (a) to (c)), Au is precipitated from the matrix, forming amorphous clusters and crystal seeds. However, not all of the initial clusters  evolve into fully developed \ce{Au} particles, as shown in frames (d) to (e). Some of these small initial clusters (and some small nuclei) are dissolved back into the disordered phase, or swallowed by other stably growing nuclei. This observation suggests the Ostwald ripening process at play, however to verify this, more statistics and larger simulation sizes are required. The final stage (f) comprises very stable AuNPs, which will continue to grow and amalgamate with each other to form large Au plates over the time scale of the simulations.
%%%%%%%%%%%%%%%%
%%%%%%%%%%%%%%%%%%%%

\begin{figure}[H]
	\centering
	\includegraphics[width=1.0\textwidth]{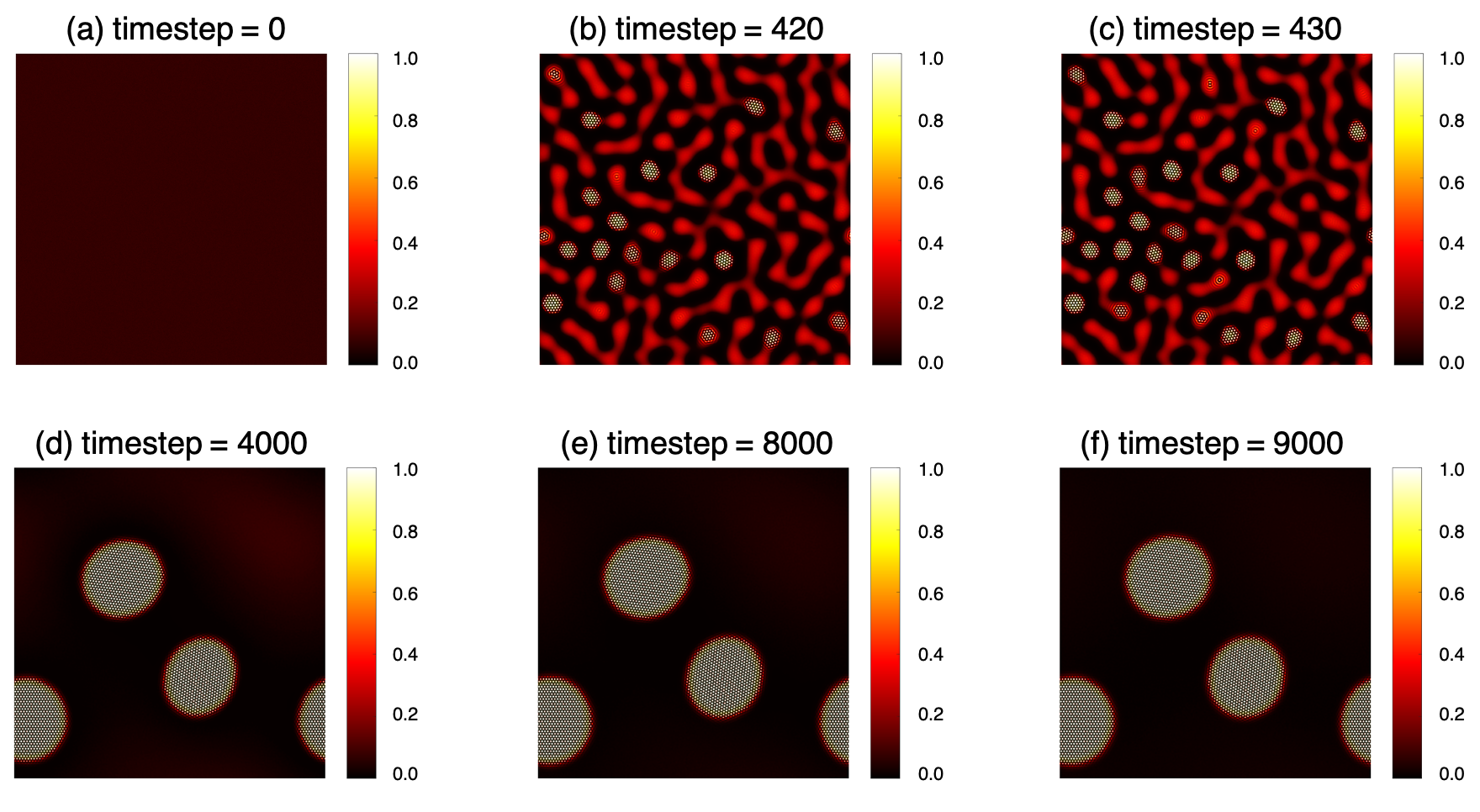}
	\caption{(Colour online) Simulated mechanosynthesis of \ce{Au}NPs from a disordered heptadecylamine (C17) -- \ce{Au(0)} matrix, with the ballistic term active ($\gamma = 0.1$ and $R = 9.7a$) for the first $430$ timesteps, and inactive for later times. The black / white pixels  represent the dimensionless density $n$ and the colourbar readings represent the composition of Au. (a-c) The initial process is the same as that Fig.~\ref{FIG:4}. (d-f) After $430$ timesteps, the ballistic term is turned off to simulate the ageing process. Here,  the rate of nucleation is nearly zero, and the initial population of nuclei under go growth and coarsening, expanding their size and decreasing in number compared to frames (d-f) in Fig.~\ref{FIG:4}.}
	\label{FIG:5}
\end{figure}

Fig.~\ref{FIG:5} shows  the result of the microstructure simulations in the Sample~3 case, where the mechanical forces is deactivated after $430$ timesteps. In frames (a) to (c) the initial stages of this simulation are the same as in Fig.~\ref{FIG:4}. In frames (d) to (f), the mechanical force is inactive and ageing occurs. In this stage, no new nuclei are created (and some of the initial nuclei also dissolve), resulting in a smaller number of nuclei growing into larger (coarser) nano-particles that in the data of Fig.~\ref{FIG:4}.
%%%%%%%%%%%%%%%%%%

Interestingly, the model revealed evolutions akin to Ostwald ripening. A more detailed illustration and statistics of the simulated mechanosysthesis are shown in the supplementary information, showing how the size distribution profiles evolve over time.
%%%%%%%%%%%%%%%%%%%

\subsection{Effect of the amine ligand size}
\label{sec:mass}
In the reference mechanosynthesis experiments  of AuNPs \cite{Rak2014}, the long-chain ligand was varied along the series pentadecylamine (C15), hexadecylamine (C16), heptadecylamine (C17), and octadecylamine (C18). The resulting AuNPs  featured an average diameter decreasing as a function of an increased chain length. In this series, the ligands vary both in length and mass. In our model, the size of the ligand is effectively described by the ligand mass variable $m_{ligand}$ and dissipation length variable $R$. Since we use an isotropic two-point correlation function to model inter-particle interactions in the XPFC model, we cannot presently capture anisotropic  atomic-scale structures in the mass density of amines. Nevertheless, we can investigate the role of chain length through the dependence of $\{m_{ligand}, R\}$ in the model, which effective captures the length scale of the ligand molecules.

\subsubsection{How the ligand size affects the kinetics and thermodynamics of the reaction}
\label{section4.2.1}

\begin{figure}[H]
	\centering
	\includegraphics[width=1.0\textwidth]{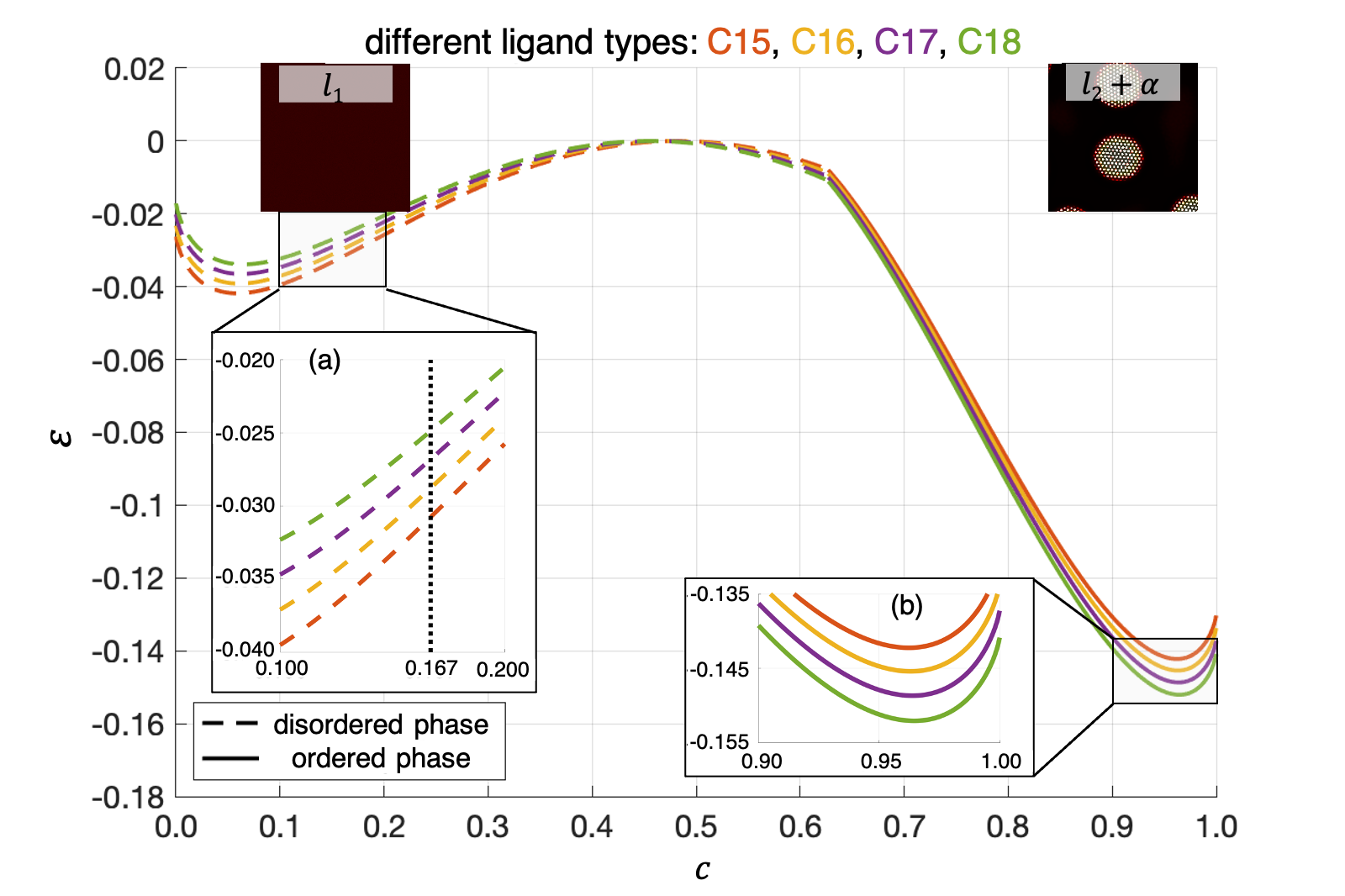}
	\caption{(Colour online) Dimensionless free energy landscape in $\mathcal{E} \times c \times\; ligand-type$ space. (Red) pentadecylamine (C15) with $m_{ligand} = 227.43 g/mol$, $R=9.1a$, (orange) hexadecylamine (C16) with $m_{ligand} = 241.46 g/mol$, $R=9.4a$, (purple) heptadecylamine (C17) with $m_{ligand} = 255.50 g/mol$, $R=9.7a$, and (green) octadecylamine (C18) with $m_{ligand} = 269.50 g/mol$, $R=10.0a$. The free energy landscapes have been subtracted from their maximum value, aligning them all at $\mathcal{E}=0$. Inset (a) shows a zoom-in of the region $c=0.1\sim 0.2$.  The black vertical dash line at $c=0.167$ marks the reaction starting point. Inset (b) shows the region of $c=0.9\sim 1.0$. The shift in each curve of the disordered phase relative to the maximum defines the dimensionless activation energy barrier $\Delta G_a$ for precipitation of of the ordered phase. }
	\label{FIG:6}
\end{figure}

Fig.~\ref{FIG:6} demonstrates the role of ligand size on the dimensionless free energy landscapes of our XPFC model, where the parameters $\{m_{ligand}, R\}$ are set accordingly for each  ligand size.
%%%%%%%%%%%%%%%%%%%%%%%%%%%
%%%%%%%%%%%%%%%%%%%%%

Fig.~\ref{FIG:6} shows that as the reactant ligand becomes longer and heavier, the free energy of the ordered phase becomes lower compared to free energy landscape's maximum, see inset Fig.~\ref{FIG:6} (b); another effect is the activation energy barrier becomes lower as ligand weight increases, as shown in inset Fig.~\ref{FIG:6} (a). For pentadecylamine (C15) the activation barrier is $\Delta G_a = 0.0309$, while for octadecylamine (C18) it is $\Delta G_a = 0.0250$, which is a reduction of of $19.11 \%$. Our model thus predicts a decreasing of the activation barrier $\Delta G_a$ with increasing %{\color{red} length and mass} 
length and mass of the ligand, which leads to a decrease in the nucleation barrier for precipitating the ordered (Au(0)) phase from the disordered ligand-rich phase. As will be shown in in section.~\ref{subsec:dyntype}, since the total number of gold atoms is constant, a consequence of this trend is that the reactions with heavier ligand produce more, but smaller crystals, a trend which is consistent with the experimental observations listed in Table. 1 of Ref.~\citenum{Rak2014}.

\subsubsection{Effect of ligand size on nanoparticle nucleation and growth morphology}
\label{subsec:dyntype}

\begin{figure}[H]
	\centering
	\includegraphics[width=1.0\textwidth]{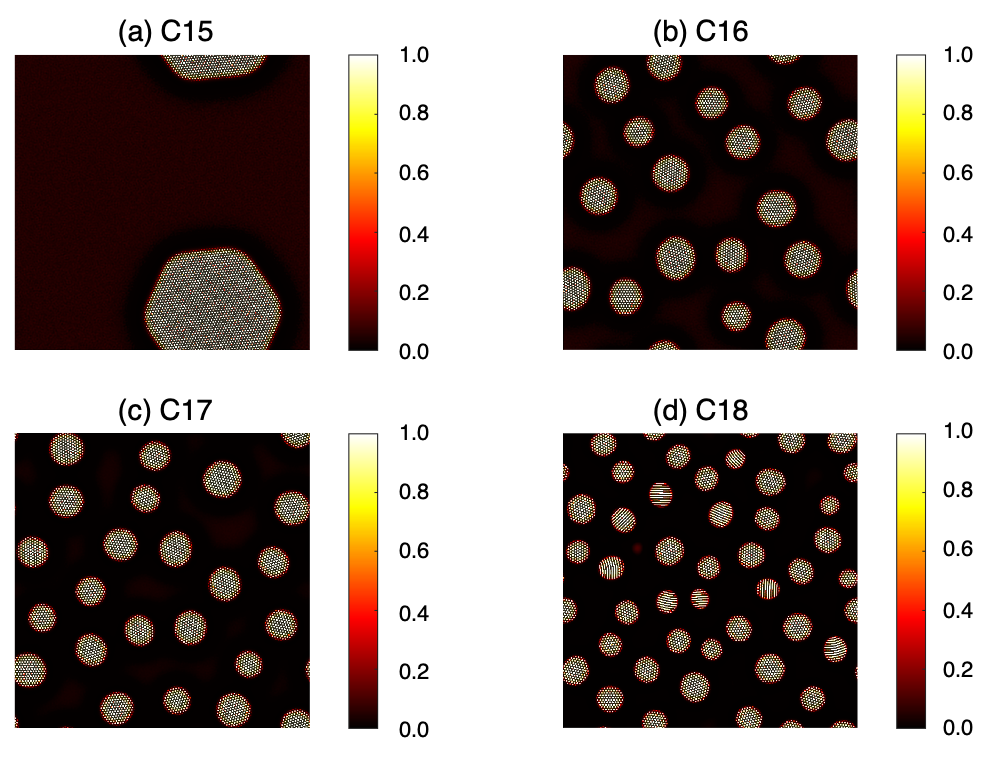}
	\caption{(Colour online) Simulated late-stage configurations of \ce{Au}NPs precipitated from a disordered ligand -\ce{Au(0)} phase. The black / white pixels represent the dimensionless density $n$, and the colourbar readings represent the composition of Au. \ce{Au}NPs for Four ligand types are shown: (a) pentadecylamine (C15), (b) hexadecylamine (C16), (c) heptadecylamine (C17), and (d) octadecylamine (C18)}
	\label{FIG:7}
\end{figure}

A second group of simulations was conducted which consisted of four samples with different ligand types and wherein the ballistic term is active with $\gamma = 0.1$, in order to simulate the reactions using \ce{HAuCl4}, and different ligands from pentadecylamine (C15) to octadecylamine (C18). The \ce{Au^{(0)}} : ligand ratio was set to be $1:5$ as in Ref.~\citenum{Rak2014}. All microstructure simulations were performed with a numerical resolution of $\Delta x = 0.125$, and the numerical timesteps were set to $\Delta t=0.00225$, $\Delta t=0.001$, $\Delta t=0.00015$, and $\Delta t=0.000015$ for pentadecylamine (C15), hexadecylamine (C16), heptadecylamine (C17), and octadecylamine (C18), respectively. Simulations were performed using a semi-implicit Fourier method described in Ref.~\citenum{Provatas2010} on a uniform two-dimensional numerical mesh of $1000 \times 1000$ two-dimensional grid points.  The average system density was set to $n_0=0.05$, while the mobility of the density and composition fields were set to $M_n = 1.0$, $M_c = 1.0$ for simplicity. The temperature was fixed at $T=0.06$. The initial configuration were once again randomly initialized for all of them. Samples~1 -- 4 correspond to the ligands, pentadecylamine (C15), hexadecylamine (C16), heptadecylamine (C17), and octadecylamine (C18), respectively. \ce{Au}NPs were found to precipitate in all samples, with some typical late-time precipitate configurations for each case shown in Fig.~\ref{FIG:7}.
%%%%%%%%%%%%%%%%%%%%
%%%%%%%%%%%%%%%%%%%%%%

The trend that emerges in the dynamical simulations of Fig.~\ref{FIG:7} is consistent with the predictions made in subsection~\ref{section4.2.1}, namely, as the ligands become longer and heavier, % {\color{red} longer and heavier}, 
the average \ce{Au}NP size becomes smaller. Table~\ref{tbl:radii} summarizes the average diameters of \ce{Au}NPs for each case and the results are compared to those in the reference experimental work.\cite{Rak2014}
%%%%%%%%%%%%%%%%%%%%%%
\begin{table}
	\caption{Simulated and experimental data of average \ce{Au}NP diameters}
	\label{tbl:radii}
	\begin{tabular}{l|ll}
		\hline
		Ligand                & Simulation (nm) & Experiment (nm) \\
		\hline
		Pentadecylamine (C15) & $22.0\pm0.0$    & $4.2\pm1.2$     \\
		Hexadecylamine (C16)  & $5.5\pm0.7$     & $1.8\pm0.3$     \\
		Heptadecylamine (C17) & $4.4\pm0.5$     & $1.5\pm0.2$     \\
		Octadecylamine (C18)  & $3.6\pm0.5$     & $1.3\pm0.2$     \\
		\hline
	\end{tabular}
\end{table}
%%%%%%%%%%%%%%%%
The trend in the average NP diameter as a function of ligand size and mass %{\color{red} size and mass} 
is the same as those observed experimentally. In particular, as the ligands become longer and heavier, %{\color{red} longer and heavier}, 
the \ce{Au}NP diameters become smaller. It is noted that the simulated \ce{Au}NP diameters are quite different {\it quantitatively} from the experimental values. This is because the XPFC model employed in this work presently uses phenomenological parameters to model the parameters of the experimental situation. As a result, the simulated diameter will be linked to these phenomenological parameters and is not expected to match the experiments quantitatively. Future works will aim to map the XPFC model onto the material properties of the experimental system, at which point a quantitative comparison will be possible.

\section{Summary and conclusions}
We have proposed a structural-phase-field-crystal (XPFC) model with a ballistic energy term to model mechanochemical reactions. The model was applied to study a mechanosynthesis experiment of ultra-small monodisperse amine-stabilized gold nanoparticles analogous to those studied experimentally in Ref.~\citenum{Rak2014}. An analysis of the kinetics of our model based on its thermodynamic free energy suggests a possible explanation of the mechanism of mechanosynthesis, namely, that the mechanical-force induced ballistic term can reduce the activation barrier energy, making the reaction (precipitated nucleation) easier to happen. A series of Langevin-type microstructure evolution simulations demonstrated the effect of the ballistic term and its role in nucleation, growth and coarsening of NPs. Using a phenomenology for representing ligand mass and length in our model, we also qualitatively reproduce the same trends for the \ce{Au}NP diameters as that observed experimentally. This provides a novel paradigm for the description of mechanochemical reaction, as well as for nanoparticles growth. While our model is internally self-consistent, the parameters of the theory on which it is based yield a phenomenological reproduction of an experimental mechanochemical system at this stage. In the future, more advanced experimental methods will need to be developed in order to provide a quantitative description of the phenomena. In particular, time resolution in experiments able to track the milling strength, frequency, density and chemical states will be needed to refine our understanding. Recently, the Nakamura group reported on tracking the nucleation of NaCl nanocrystals inside a carbon nanotube, resolved in time, by transmission electron microscopy\cite{Nakamuro2021}, opening opportunities to obtain quantitative data to back models such as ours. 

\begin{acknowledgement}

	All the authors thank the Natural Science and Engineering Research Council of Canada (NSERC) Discovery Grant and Discovery Accelerator Supplement programs, the Canada Research Chairs (CRC) Program, the McGill Sustainability Systems Initiative (MSSI) and McGill University for their financial support. We are grateful to Nana Ofori-Opoku for help with consolidating the structural-phase-field-crystal model with ballistic term, we thank Matthew Frick and Paul Jreidini for the discussions on the models. Monika Rak, Blaine Fiss, Boris Nijikovsky and Michael Ferguson are acknowledged for fruitful discussions on nanoparticle mechanosynthesis.

\end{acknowledgement}

\newpage

\begin{suppinfo}

\counterwithin{figure}{section}
\counterwithin*{equation}{section}
\counterwithin*{equation}{subsection}
\renewcommand{\thefigure}{S\arabic{figure}}

\section{Supporting Information}

\subsection{XPFC Model for Mechanosynthesis: detailed derivations}
	\label{sec:modelexp}

	The detailed derivation of the new XPFC model in section 2 is shown in this supplementary information section.

	We begin with a dimensionless generalized binary XPFC free energy functional \cite{Smith2017}
	\begin{align}
		\frac{\Delta\mathcal{F}[n,c]}{k_{B}T\rho_{0}} = \frac{\Delta\mathcal{F}_{id}[n]}{k_{B}T\rho_{0}} + \frac{\Delta\mathcal{F}_{mix}[n,c]}{k_{B}T\rho_{0}} + \frac{\Delta\mathcal{F}_{ex}[n,c]}{k_{B}T\rho_{0}},
	\end{align}
	where $k_{B}$ is Boltzmann's constant, $T$ is the temperature, and $\rho_{0}$ is the total system reference density. The free energy functional has been broken into the ideal $\Delta\mathcal{F}_{id}$, mixing $\Delta\mathcal{F}_{mix}$, and excess $\Delta\mathcal{F}_{ex}$ parts, respectively. The field $n$,  defined as $n=(\rho-\rho_0)/\rho_0$ where $\rho$ is the local mass density, plays the role of an order parameter of the model. By adopting a series of reasonable approximations for each term, the free energy form is developed in terms of $n$.

	The ideal part of the free energy is defined as
	\begin{align}
		\frac{\Delta\mathcal{F}_{id}[n]}{k_{B}T\rho_{0}} = \int d\vec{x}\left\{\, \left[n(\vec{x})+1\right]\ln\left[n(\vec{x})+1\right] - n(\vec{x})\, \right\}
	\end{align}
	Following \citeauthor{PhysRevB.75.064107}, we simplify this term by expanding it about its reference density in a polynomial truncated to fourth order \cite{PhysRevB.75.064107}, yielding
	\begin{align}
		\frac{\Delta\mathcal{F}_{id}[n]}{k_{B}T\rho_{0}} = \int d\vec{x} \left\{\frac{n(\vec{x})^2}{2}-\eta\frac{n(\vec{x})^3}{6} + \chi\frac{n(\vec{x})^4}{12}\right\},
	\end{align}
	where the phenomenological parameters $\eta$ and $\chi$ are added to fit the free energy away from the reference density.

	The mixing part of the free energy is taken as
	\begin{align}
		\frac{\Delta\mathcal{F}_{mix}[n,c]}{k_{B}T\rho_{0}} = \int d\vec{x} \left\{\left[n(\vec{x})+1\right]\cdot\left[c \, \ln\left(\frac{c}{c_{0}}\right) + \left(1-c\right)\, \ln\left(\frac{1-c}{1-c_{0}}\right)\right]\right\},
	\end{align}
	where $c_{0}$ is the composition of the reference state. This follows the original free energy of an ideal alloy of \citeauthor{Ofori-Opoku2012} \cite{Ofori-Opoku2012}. \citeauthor{Smith2017} generalized the mixing term to include enthalpy \cite{Smith2017} as follows,
	\begin{align}
		\frac{\Delta\mathcal{F}_{mix}[n,c]}{k_{B}T\rho_{0}} = \int d\vec{x} \,\, \omega \left\{\left[n(\vec{x})+1\right]\cdot\left[c \, \ln\left(\frac{c}{c_{0}}\right) + \left(1-c\right)\, \ln\left(\frac{1-c}{1-c_{0}}\right)\right] + \frac{1}{2}\epsilon(c-c_{0})^2 \right\},
		\label{eq:fmix}
	\end{align}
	where the phenomenological parameter $\omega$ is added to modify the mixing term away from the reference states, and the $({1}/{2})\epsilon(c-c_{0})^2$ term is added to account for the enthalpy of mixing, where $\epsilon$ is given by
	\begin{align}
		\epsilon = -4 + \epsilon_{0}(T-T_{c})
	\end{align}
	In the spirit of Landau theory, the enthalpy term has been truncated to quadratic order in $c$, with a coefficient linearly proportional to temperature $T$.

	The excess part of the free energy is responsible for capturing inter-particle interactions. It is defined as
	%%%%%%%%%%%%%%%%%
	\begin{align}
		\frac{\Delta\mathcal{F}_{ex}[n,c]}{k_{B}T\rho_{0}} & = -\frac{1}{2} \left\{ n(\vec{x}) * C_{nn}(\vec{x}, \vec{x}') * n(\vec{x}') \right. \nonumber \\
		                                                                          & + n(\vec{x}) * C_{nc}(\vec{x}, \vec{x}') * \Delta c(\vec{x}') \nonumber                       \\
		                                                                          & + \Delta c(\vec{x}) * C_{cn}(\vec{x}, \vec{x}') * n(\vec{x}') \nonumber                       \\
		                                                                          & + \left. \Delta c(\vec{x}) * C_{cc}(\vec{x}, \vec{x}') * \Delta c(\vec{x}') \right\},
	\end{align}
	%%%%%%%%%%%%%%%%%
	where $\Delta c(\vec{x})=c(\vec{x})-c_{0}$ is the composition difference from the reference and $*$ is the convolution operation. \citeauthor{Greenwood2011} further simplify this term by noting that since the microscopic density $n$ oscillates rapidly, while $c$ is a smooth field in the long wavelength limit (i.e. on scales larger than the lattice constant), the cross terms of $n$ and $c$ vanish, leaving only the quadratic terms. Approximating $C_{cc}$ by a square gradient composition contribution as in  Cahn-Hilliard theory thus \cite{Greenwood2011} gives
	\begin{align}
		\frac{\Delta\mathcal{F}_{ex}[n,c]}{k_{B}T\rho_{0}} & = -\frac{1}{2}\int d\vec{x} \left\{ n(\vec{x}) \int d\vec{x}'  C_{nn}(\vec{x}, \vec{x}') \cdot n(\vec{x}') -W_{c} |\nabla c(\vec{x})|^2\right\},
	\end{align}
	where the phenomenological parameter $W_{c}$, is introduced to describe the energy costs associated with compositional interfaces. The form of $C_{nn}$ follows the work of \citeauthor{Smith2017} who simply decomposed $C_{nn}$ as a summation of the corresponding correlation functions of each component of the system modulated by the local composition $c(\vec{x})$ \cite{Smith2017}. This is done through the use of an interpolating function denoted by $\zeta(c)$, and allows the freedom of the system to select crystal structures depending on the local solute composition. The specific form of $C_{nn}$ in this phenomenology is given by
	\begin{align}
		C_{eff} \coloneqq C_{nn} = \zeta_{A}(1-c)\cdot C_{A} + \zeta_{B}(c)\cdot C_{B},
		\label{eq:Ceff}
	\end{align}
	where ${\zeta_A}/{\zeta_B}=1$ or $0$ when $c=0$ or $1$. Specific forms of the $\zeta_i$ functions ($i=A$ or $B$) are discussed further below. In Eq.~(\ref{eq:Ceff}), $C_A$ and $C_B$ are the direct density-density correlation functions used to model interactions in the pure components $i=A$ or $B$, respectively. Their form follows previous works~\cite{PhysRevB.75.064107, Greenwood2011, Smith2017},  written in Fourier space as
	\begin{align}
		\hat{C}_i=e^{-T/T_i}\sum_{j}e^{-\frac{(k-k_j^i)^2}{2\sigma_j^2}},
	\end{align}
	where $\hat{C}_i$ denotes the Fourier transform of $C_i$ and $j$ indexes the primary reflection peak of the $j^{th}$ family of lattice planes of the crystal structure adopted by component $i$. Here, $k_j^i$ is the lattice spacing of the $j^{th}$ family of planes of component $i$. The peaks of the correlation functions of each component are modulated by an exponential in temperature ($T$), and $T_i$ is a parameter to be set such as to activate crystallization at an appropriate temperature in each component. As shown by \citeauthor{Smith2017} this system can be well approximated by choosing
	\begin{align}
		\zeta_{A}(1-c) & = 0 \nonumber                         \\
		\zeta_{B}(c)   & = e^{-\frac{(1-c)^2}{2\alpha_{c}^2}},
	\end{align}
	%%%%%
	which yields the following effective two-point density correlation function,
	%%%%%%%
	\begin{align}
		\hat{C}_{eff} = e^{-T/T_{B}}e^{-\frac{(1-c)^2}{2\alpha_{c}^2}}e^{-\frac{(k-k_{B})^2}{2\sigma^2}}
		\label{eq:correffect}
	\end{align}
	%%%%%%%%
	Equation~(\ref{eq:correffect}) is used to model inter-particle interactions in the monotectic mixture studied in the remainder of this work. In which, $\alpha_{c}$ is the peak width of the interpolating function $\zeta_{B}$, which also serves as a ``window function'' describing how the system behaves like a pure B system, $k_{B}=\frac{2\pi}{a_{B}}$ is the peak of the correlation function for pure B system, assuming that B forms a hexagonal crystal with a lattice constant of $a_{B}$. It is noted that there should strictly be a $k=0$ component of $\hat{C}_{eff}$. For simplicity, it is left out of this work.

	\subsection{Weight function $w_{R}(x)$}
	\label{sec:weight}

	The weight function $w_{R}(x)$ describes how the external mechanical energy dissipates across the system, which can be modulated in a variety of ways, for example, in \citeauthor{Enrique2000}'s Ref.~\citenum{Enrique2000}, Yukawa potential is used, in \citeauthor{Ofori-Opoku2012}'s Ref.~\citenum{Ofori-Opoku2012}, Bessel function of the second kind is chosen. In this model we adopted a normalized Gaussian function,
	%%%%%%%%%%
	\begin{align}
		w_{R}\left(x \coloneqq |\vec{x} - \vec{x}'|\right) = \frac{1}{R\sqrt{2\pi}}e^{-\frac{1}{2}\left(\frac{x}{R}\right)^2},
	\end{align}
	%%%%%%%%%%
	where $R$ is the energy dissipation length. We fixed $R$ to a value that scaled with the ligand chain length. We do not know this scaling factor at the present time, as a result, we estimate $R = 0.75\; \times $ chain length $+\; 4.825$. In this work, we study 4 samples pentadecylamine (denoted C15), hexadecylamine (denoted C16), heptadecylamine (denoted C17), and octadecylamine (denoted C18), with chain lengths equal to  5.7a, 6.1a, 6.5a, 6.9a (a is the lattice constant of Au), respectively. For these samples, the values of R then become $R = 9.1,\; 9.4,\; 9.7,\; 10.0$, respectively.

	\subsection{Equilibrium properties: detailed derivations}
	\label{sec:equilibrium_propsexp}

	The detailed derivation of the equilibrium bulk properties of this model is shown in this supplementary information section.

	To investigate the equilibrium properties of the model in Eq.~(11) in main paper, we describe the bulk ordered states by a mode-expansion for the density around an average density,
	%%%%%%%%%%%%%%%%%
	\begin{align}
		n(\vec{x}) = n_0(\vec{x}) + \sum_{\vec{G}_{j}}\phi_{j}(\vec{x})e^{i\vec{G}_{j}\cdot\vec{x}} + c.c.
		\label{mode_expansion}
	\end{align}
	%%%%%%%%%%%%%%%%%
	where $c.c.$ is is the complex conjugate, $n_0(\vec{x})$ is the average density of the system, $\vec{G}_{j}$ represents the reciprocal basis vector of mode $j$, and $\phi_{j}$ represents the amplitude of mode $j$. For the equilibrium analysis it is further assumed that all modes describing the crystal of interest have the same real amplitude, making $\phi_j=\phi$ for all $j$. it is noted that in the context of Eq.~(\ref{mode_expansion}), disordered  phases are described by $\phi_j=0$, and $n_o(\vec{x})$ becomes a constant. Thus, for a crystalline phase hexagonal symmetry, such as gold nano-particles in 2D, the density expression becomes
	%%%%%%%%%%%%%%%%%%%
	\begin{align}
		n(\vec{x}) = n_0(\vec{x}) + 2\phi\left\{ \cos\left[\frac{2\pi}{a}\left(\frac{2y}{\sqrt{3}}\right)\right] + 4 \cos\left(\frac{2\pi x}{a}\right)\cos\left[\frac{2\pi}{a}\left(\frac{2y}{\sqrt{3}}\right)\right]\,\right\}
		\label{eq:densityexp}
	\end{align}
	%%%%%%%%%%%%%%%%%

	We next proceed by substituting Eq.~\ref{eq:densityexp} into Eq.~(11) in main paper and integrating over the area (in 2D) of a unit cell of the crystal described by the expansion in Eq.(\ref{eq:densityexp}). In so doing, the exponential terms can be integrated exactly (with some vanishing and some giving non-zero contributions). This process is a simple form of coarse graining and is described operationally in detail in Ref.~\citenum{Provatas2010}. The resulting coarse grained free energy is interpreted as the free energy in terms of $\phi$ and reduces to the bulk disordered phase when $\phi=0$ and that of the bulk solid when $\phi \ne 0$. In the former case it is a function of the average (bulk) density $n_0$ and composition $c$, while in the latter case it is a function of $n_0$, the composition $c$ {\it and} the order parameter $\phi$. In scaled units ($\mathcal{E} = {\Delta\mathcal{F}}/{k_{B}T\rho_{0}V}$) the explicit form of the bulk free energy of the disordered phase (denoted $\mathcal{E}_{d}$) and the ordered hexagonal phase (denoted $\mathcal{E}_{o}$) become Eq.~(12) and Eq.~(13) in main paper, respectively.

	\subsection{Dynamics of the structural-phase-field-crystal model}
	We assume that in our model, total density $n$ and total composition $c$ are all conserved, the dynamics of the structural-phase-field-crystal model follow Langevin type conserved equations:
	%%%%%%%%%%%%%%%%%%%%
	\begin{align}
		\frac{\partial n}{\partial t} & = M_{n}\nabla^2\left( \frac{\delta \mathcal{F}}{\delta n}\right) + \vec{\nabla}\cdot \vec{\xi}_{n} \nonumber \\
		\frac{\partial c}{\partial t} & = M_{c}\nabla^2\left( \frac{\delta \mathcal{F}}{\delta c}\right) + \vec{\nabla}\cdot \vec{\xi}_{c}
		\label{eq:dynamics}
	\end{align}
	%%%%%%%%%%%%%%%
	where $\mathcal{F} = {\Delta\mathcal{F}[n,c]}/{k_{B}T\rho_{0}}$ is the reduced dimensionless free energy, while $\vec{\xi}_{n}$ and $\vec{\xi}_{c}$ are introduced to model the effect of thermal fluctuations on wavelengths larger than the atomic scale, which are necessary provide the fluctuations from which nucleation can arise spontaneously. From statistical thermodynamics, it can be shown that the noise sources must satisfy the  fluctuation-dissipation relation:
	%%%%%%%%%%%%%%%%%%
	\begin{align}
		\langle\vec{\xi}_{n/c}^{j}(\vec{x}, t),\vec{\xi}_{n/c}^{i}(\vec{x}', t')\rangle = -2M_{n/c}\delta(\vec{x}-\vec{x}')\delta(t-t')\delta_{ij}
		\label{eq:noise}
	\end{align}

	\subsection{Simulated mechanosynthesis of AuNPs}

	In this section we provide a more detailed simulated results of the mechanosynthesis of AuNPs from a disordered heptadecylamine (C17) - Au(0) matrix, $\{n, c\}$ data of Sample. 2 and Sample. 3 (defined in main paper subsection 4.1.2) are visualized, the corresponding histogram of the NP diameters are calculated. Fig.~\ref{FIG:S1}, Fig.~\ref{FIG:S2}, and Fig.~\ref{FIG:S3} visualize the NP morphology evolving over timesteps:
	\begin{itemize}
		\item Once there is no mechanical energy, there would be no new nuclei embryos formation. Comparing from inset (c) to (d.1) and (d.2) in Fig.~\ref{FIG:S1}: At timestep $430$, there is already $11$ embryos as shown in inset (c). At timestep $550$, the number of embryos grow to $52$ with the ballistic term continuously active as shown in inset (d.1); however, that number decreases to $3$ once the ballistic term is deactivated as shown in inset (d.2).
		\item Ostwald ripening is observed to  happen during the reaction. We present two cases during the simulation. The first case is shown in Fig.~\ref{FIG:S2}, from inset (b. 1) to (d. 1) (timestep from $1600$ to $2400$): tiny nuclei embryos resolve, remaining embryos grow larger and larger, in the histograms, the ripening can be observed by the fact that diameter $< 3nm$ counts appear (embryos shrink), then disappear (embryos resolve). The second case is shown in Fig.~\ref{FIG:S3}, from inset (a. 1) to (c. 1) (timestep from $3500$ to $4500$); again, the tiny embryos with diameter $< 3nm$ resolve, remaining embryos form nuclei, then grow to NPs.
	\end{itemize}

	\newpage
	\begin{figure}[H]
		\centering
		\includegraphics[width=1.0\textwidth]{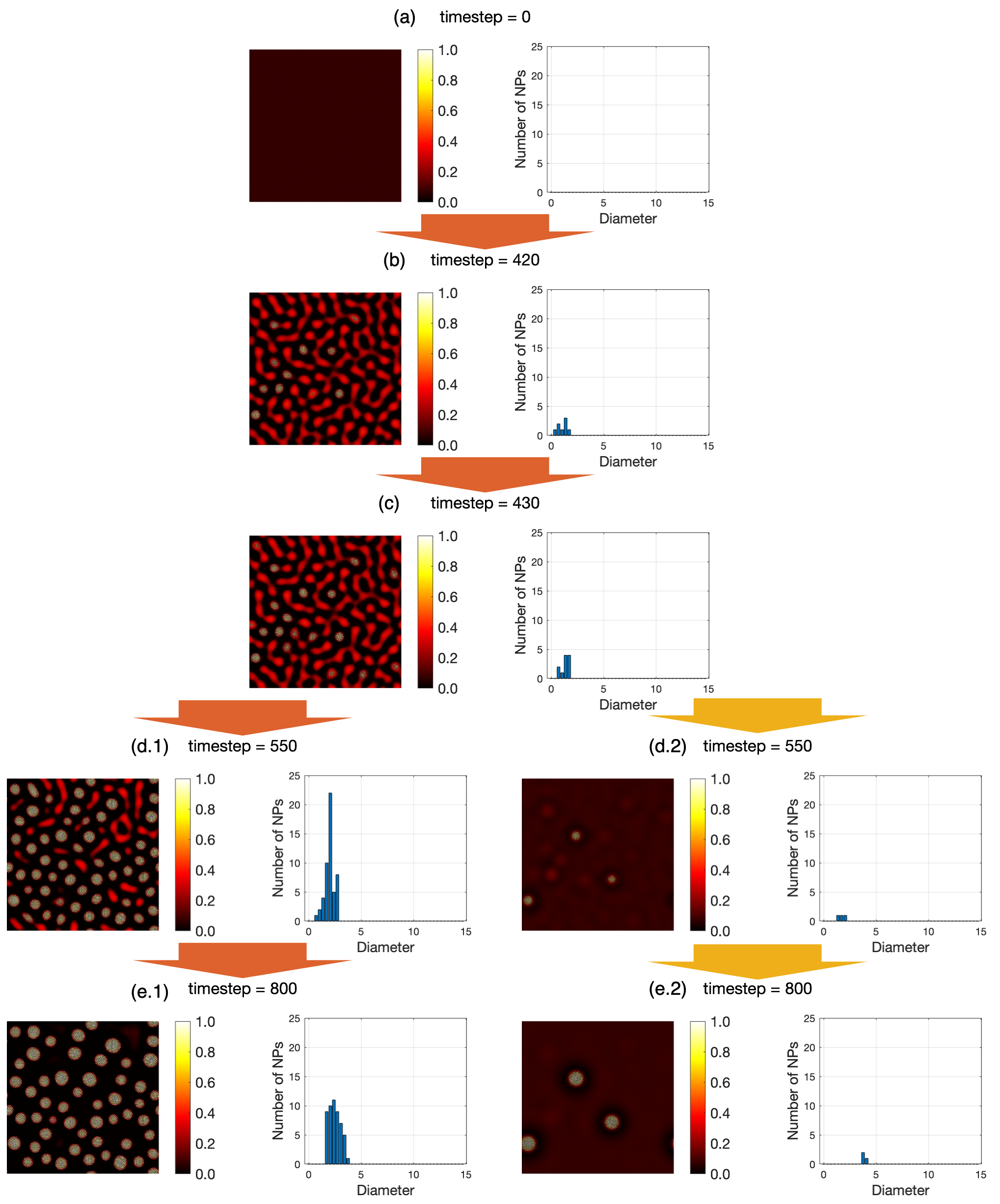}
		\caption{(Colour online) Simulated mechanosynthesis of \ce{Au}NPs from a disordered heptadecylamine (C17) -- \ce{Au(0)} matrix. The black / white pixels represent the dimensionless density $n$ and the colourbar readings represent the composition of Au. (a-c), (d.1) and (e.1) are with the ballistic term continuously active ($\gamma = 0.1$). (d.2) and (e.2) are with the ballistic term deactivated.}
		\label{FIG:S1}
	\end{figure}

	\newpage
	\begin{figure}[H]
		\centering
		\includegraphics[width=1.0\textwidth]{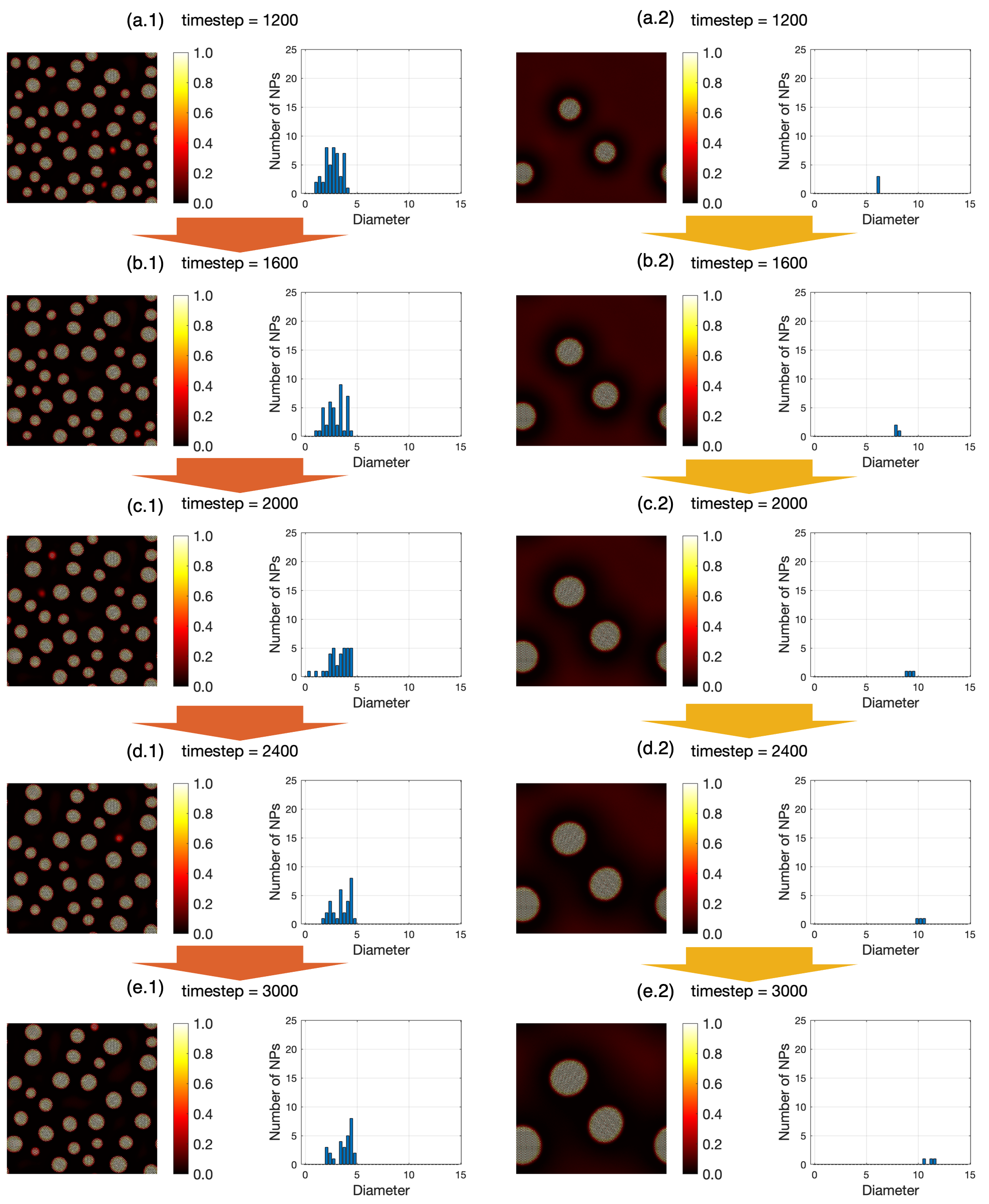}
		\caption{(Colour online) Simulated mechanosynthesis of \ce{Au}NPs from a disordered heptadecylamine (C17) -- \ce{Au(0)} matrix (continued). The black / white pixels represent the dimensionless density $n$ and the colourbar readings represent the composition of Au. (a-e.1) are with the ballistic term continuously active ($\gamma = 0.1$). (a-e.2) are with the ballistic term deactivated.}
		\label{FIG:S2}
	\end{figure}

	\newpage
	\begin{figure}[H]
		\centering
		\includegraphics[width=1.0\textwidth]{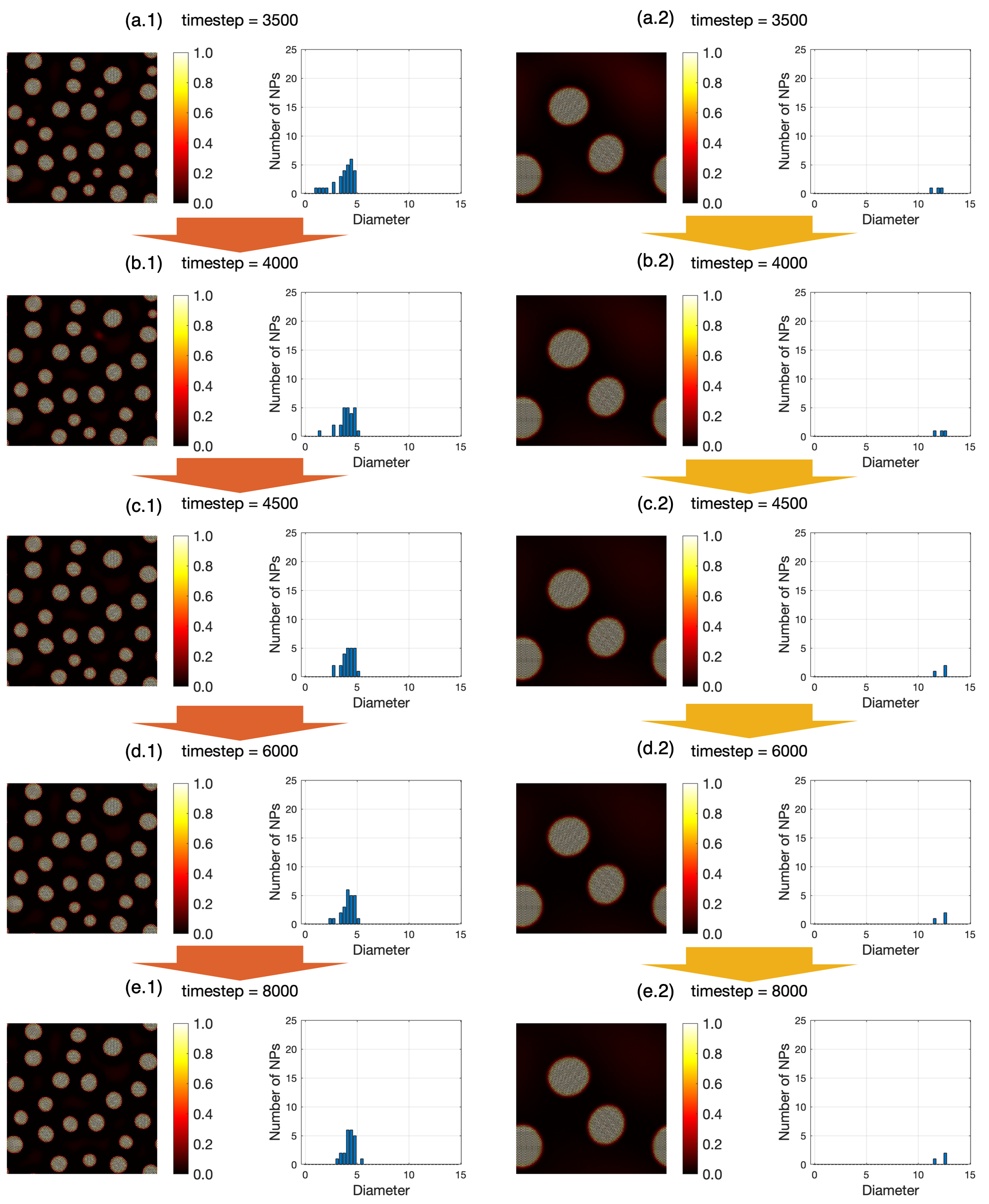}
		\caption{(Colour online) Simulated mechanosynthesis of \ce{Au}NPs from a disordered heptadecylamine (C17) -- \ce{Au(0)} matrix (continued). The black / white pixels represent the dimensionless density $n$ and the colourbar readings represent the composition of Au. (a-e.1) are with the ballistic term continuously active ($\gamma = 0.1$). (a-e.2) are with the ballistic term deactivated.}
		\label{FIG:S3}
	\end{figure}

\end{suppinfo}

\bibliography{manuscript}

\providecommand{\latin}[1]{#1}
\makeatletter
\providecommand{\doi}
  {\begingroup\let\do\@makeother\dospecials
  \catcode`\{=1 \catcode`\}=2 \doi@aux}
\providecommand{\doi@aux}[1]{\endgroup\texttt{#1}}
\makeatother
\providecommand*\mcitethebibliography{\thebibliography}
\csname @ifundefined\endcsname{endmcitethebibliography}
  {\let\endmcitethebibliography\endthebibliography}{}
\begin{mcitethebibliography}{80}
\providecommand*\natexlab[1]{#1}
\providecommand*\mciteSetBstSublistMode[1]{}
\providecommand*\mciteSetBstMaxWidthForm[2]{}
\providecommand*\mciteBstWouldAddEndPuncttrue
  {\def\EndOfBibitem{\unskip.}}
\providecommand*\mciteBstWouldAddEndPunctfalse
  {\let\EndOfBibitem\relax}
\providecommand*\mciteSetBstMidEndSepPunct[3]{}
\providecommand*\mciteSetBstSublistLabelBeginEnd[3]{}
\providecommand*\EndOfBibitem{}
\mciteSetBstSublistMode{f}
\mciteSetBstMaxWidthForm{subitem}{(\alph{mcitesubitemcount})}
\mciteSetBstSublistLabelBeginEnd
  {\mcitemaxwidthsubitemform\space}
  {\relax}
  {\relax}

\bibitem[James \latin{et~al.}(2012)James, Adams, Bolm, Braga, Collier,
  Fri{\v{s}}{\v{c}}i{\'c}, Grepioni, Harris, Hyett, Jones, Krebs, Mack, Maini,
  Orpen, Parkin, Shearouse, Steed, and Waddell]{james2012mechanochemistry}
James,~S.~L.; Adams,~C.~J.; Bolm,~C.; Braga,~D.; Collier,~P.;
  Fri{\v{s}}{\v{c}}i{\'c},~T.; Grepioni,~F.; Harris,~K. D.~M.; Hyett,~G.;
  Jones,~W.; Krebs,~A.; Mack,~J.; Maini,~L.; Orpen,~A.~G.; Parkin,~I.~P.;
  Shearouse,~W.~C.; Steed,~J.~W.; Waddell,~D.~C. Mechanochemistry:
  Opportunities for New and Cleaner Synthesis. \emph{Chemical Society Reviews}
  \textbf{2012}, \emph{41}, 413--447\relax
\mciteBstWouldAddEndPuncttrue
\mciteSetBstMidEndSepPunct{\mcitedefaultmidpunct}
{\mcitedefaultendpunct}{\mcitedefaultseppunct}\relax
\EndOfBibitem
\bibitem[Takacs(2013)]{C2CS35442J}
Takacs,~L. The Historical Development of Mechanochemistry. \emph{Chem. Soc.
  Rev.} \textbf{2013}, \emph{42}, 7649--7659\relax
\mciteBstWouldAddEndPuncttrue
\mciteSetBstMidEndSepPunct{\mcitedefaultmidpunct}
{\mcitedefaultendpunct}{\mcitedefaultseppunct}\relax
\EndOfBibitem
\bibitem[Hernández and Bolm(2017)Hernández, and
  Bolm]{doi:10.1021/acs.joc.6b02887}
Hernández,~J.~G.; Bolm,~C. Altering Product Selectivity by Mechanochemistry.
  \emph{The Journal of Organic Chemistry} \textbf{2017}, \emph{82},
  4007--4019\relax
\mciteBstWouldAddEndPuncttrue
\mciteSetBstMidEndSepPunct{\mcitedefaultmidpunct}
{\mcitedefaultendpunct}{\mcitedefaultseppunct}\relax
\EndOfBibitem
\bibitem[Andersen and Mack(2018)Andersen, and Mack]{C7GC03797J}
Andersen,~J.; Mack,~J. Mechanochemistry and Organic Synthesis: from Mystical to
  Practical. \emph{Green Chem.} \textbf{2018}, \emph{20}, 1435--1443\relax
\mciteBstWouldAddEndPuncttrue
\mciteSetBstMidEndSepPunct{\mcitedefaultmidpunct}
{\mcitedefaultendpunct}{\mcitedefaultseppunct}\relax
\EndOfBibitem
\bibitem[Howard \latin{et~al.}(2018)Howard, Cao, and Browne]{C7SC05371A}
Howard,~J.~L.; Cao,~Q.; Browne,~D.~L. Mechanochemistry as an Emerging Tool for
  Molecular Synthesis: What Can It Offer? \emph{Chem. Sci.} \textbf{2018},
  \emph{9}, 3080--3094\relax
\mciteBstWouldAddEndPuncttrue
\mciteSetBstMidEndSepPunct{\mcitedefaultmidpunct}
{\mcitedefaultendpunct}{\mcitedefaultseppunct}\relax
\EndOfBibitem
\bibitem[Bonnamour \latin{et~al.}(2013)Bonnamour, Métro, Martinez, and
  Lamaty]{C3GC40302E}
Bonnamour,~J.; Métro,~T.-X.; Martinez,~J.; Lamaty,~F. Environmentally Benign
  Peptide Synthesis Using Liquid-assisted Ball-milling: Application to the
  Synthesis of Leu-enkephalin. \emph{Green Chem.} \textbf{2013}, \emph{15},
  1116--1120\relax
\mciteBstWouldAddEndPuncttrue
\mciteSetBstMidEndSepPunct{\mcitedefaultmidpunct}
{\mcitedefaultendpunct}{\mcitedefaultseppunct}\relax
\EndOfBibitem
\bibitem[Tan \latin{et~al.}(2016)Tan, Loots, and Friščić]{C6CC02015A}
Tan,~D.; Loots,~L.; Friščić,~T. Towards Medicinal Mechanochemistry:
  Evolution of Milling from Pharmaceutical Solid Form Screening to the
  Synthesis of Active Pharmaceutical Ingredients (APIs). \emph{Chem. Commun.}
  \textbf{2016}, \emph{52}, 7760--7781\relax
\mciteBstWouldAddEndPuncttrue
\mciteSetBstMidEndSepPunct{\mcitedefaultmidpunct}
{\mcitedefaultendpunct}{\mcitedefaultseppunct}\relax
\EndOfBibitem
\bibitem[Mottillo and Fri{\v{s}}{\v{c}}i{\'c}(2017)Mottillo, and
  Fri{\v{s}}{\v{c}}i{\'c}]{mottillo2017advances}
Mottillo,~C.; Fri{\v{s}}{\v{c}}i{\'c},~T. Advances in Solid-state
  Transformations of Coordination Bonds: from the Ball Mill to the Aging
  Chamber. \emph{Molecules} \textbf{2017}, \emph{22}, 144\relax
\mciteBstWouldAddEndPuncttrue
\mciteSetBstMidEndSepPunct{\mcitedefaultmidpunct}
{\mcitedefaultendpunct}{\mcitedefaultseppunct}\relax
\EndOfBibitem
\bibitem[Quaresma \latin{et~al.}(2017)Quaresma, Andr{\'e}, Fernandes, and
  Duarte]{quaresma2017mechanochemistry}
Quaresma,~S.; Andr{\'e},~V.; Fernandes,~A.; Duarte,~M.~T. Mechanochemistry--a
  Green Synthetic Methodology Leading to Metallodrugs, Metallopharmaceuticals
  and Bio-inspired Metal-organic Frameworks. \emph{Inorganica Chimica Acta}
  \textbf{2017}, \emph{455}, 309--318\relax
\mciteBstWouldAddEndPuncttrue
\mciteSetBstMidEndSepPunct{\mcitedefaultmidpunct}
{\mcitedefaultendpunct}{\mcitedefaultseppunct}\relax
\EndOfBibitem
\bibitem[Fri{\v{s}}{\v{c}}i{\'{c}}
  \latin{et~al.}(2013)Fri{\v{s}}{\v{c}}i{\'{c}}, Halasz, Beldon, Belenguer,
  Adams, Kimber, Honkim{\"{a}}ki, and Dinnebier]{Friscic2013}
Fri{\v{s}}{\v{c}}i{\'{c}},~T.; Halasz,~I.; Beldon,~P.~J.; Belenguer,~A.~M.;
  Adams,~F.; Kimber,~S.~A.; Honkim{\"{a}}ki,~V.; Dinnebier,~R.~E. {Real-time
  and in situ Monitoring of Mechanochemical Milling Reactions}. \emph{Nature
  Chemistry} \textbf{2013}, \emph{5}, 66--73\relax
\mciteBstWouldAddEndPuncttrue
\mciteSetBstMidEndSepPunct{\mcitedefaultmidpunct}
{\mcitedefaultendpunct}{\mcitedefaultseppunct}\relax
\EndOfBibitem
\bibitem[Julien \latin{et~al.}(2017)Julien, Mottillo, and
  Friščić]{C7GC01078H}
Julien,~P.~A.; Mottillo,~C.; Friščić,~T. Metal–organic Frameworks Meet
  Scalable and Sustainable Synthesis. \emph{Green Chem.} \textbf{2017},
  \emph{19}, 2729--2747\relax
\mciteBstWouldAddEndPuncttrue
\mciteSetBstMidEndSepPunct{\mcitedefaultmidpunct}
{\mcitedefaultendpunct}{\mcitedefaultseppunct}\relax
\EndOfBibitem
\bibitem[Fidelli \latin{et~al.}(2018)Fidelli, Karadeniz, Howarth, Huskić,
  Germann, Halasz, Etter, Moon, Dinnebier, Stilinović, Farha, Friščić, and
  Užarević]{C8CC03189D}
Fidelli,~A.~M.; Karadeniz,~B.; Howarth,~A.~J.; Huskić,~I.; Germann,~L.~S.;
  Halasz,~I.; Etter,~M.; Moon,~S.-Y.; Dinnebier,~R.~E.; Stilinović,~V.;
  Farha,~O.~K.; Friščić,~T.; Užarević,~K. Green and Rapid Mechanosynthesis
  of High-porosity NU- and UiO-type Metal–organic Frameworks. \emph{Chem.
  Commun.} \textbf{2018}, \emph{54}, 6999--7002\relax
\mciteBstWouldAddEndPuncttrue
\mciteSetBstMidEndSepPunct{\mcitedefaultmidpunct}
{\mcitedefaultendpunct}{\mcitedefaultseppunct}\relax
\EndOfBibitem
\bibitem[Stolar and Užarević(2020)Stolar, and Užarević]{D0CE00091D}
Stolar,~T.; Užarević,~K. Mechanochemistry: an Efficient and Versatile Toolbox
  for Synthesis{,} Transformation{,} and Functionalization of Porous
  Metal–organic Frameworks. \emph{CrystEngComm} \textbf{2020}, \emph{22},
  4511--4525\relax
\mciteBstWouldAddEndPuncttrue
\mciteSetBstMidEndSepPunct{\mcitedefaultmidpunct}
{\mcitedefaultendpunct}{\mcitedefaultseppunct}\relax
\EndOfBibitem
\bibitem[Garay \latin{et~al.}(2007)Garay, Pichon, and James]{B600363J}
Garay,~A.~L.; Pichon,~A.; James,~S.~L. Solvent-free Synthesis of Metal
  Complexes. \emph{Chem. Soc. Rev.} \textbf{2007}, \emph{36}, 846--855\relax
\mciteBstWouldAddEndPuncttrue
\mciteSetBstMidEndSepPunct{\mcitedefaultmidpunct}
{\mcitedefaultendpunct}{\mcitedefaultseppunct}\relax
\EndOfBibitem
\bibitem[Prochowicz \latin{et~al.}(2017)Prochowicz, Yadav, Saliba, Saski,
  Zakeeruddin, Lewiński, and Grätzel]{C7SE00094D}
Prochowicz,~D.; Yadav,~P.; Saliba,~M.; Saski,~M.; Zakeeruddin,~S.~M.;
  Lewiński,~J.; Grätzel,~M. Mechanosynthesis of Pure Phase Mixed-cation
  MAxFA1−xPbI3 Hybrid Perovskites: Photovoltaic Performance and
  Electrochemical Properties. \emph{Sustainable Energy Fuels} \textbf{2017},
  \emph{1}, 689--693\relax
\mciteBstWouldAddEndPuncttrue
\mciteSetBstMidEndSepPunct{\mcitedefaultmidpunct}
{\mcitedefaultendpunct}{\mcitedefaultseppunct}\relax
\EndOfBibitem
\bibitem[Prochowicz \latin{et~al.}(2019)Prochowicz, Saski, Yadav, Grätzel, and
  Lewiński]{doi:10.1021/acs.accounts.9b00454}
Prochowicz,~D.; Saski,~M.; Yadav,~P.; Grätzel,~M.; Lewiński,~J.
  Mechanoperovskites for Photovoltaic Applications: Preparation,
  Characterization, and Device Fabrication. \emph{Accounts of Chemical
  Research} \textbf{2019}, \emph{52}, 3233--3243\relax
\mciteBstWouldAddEndPuncttrue
\mciteSetBstMidEndSepPunct{\mcitedefaultmidpunct}
{\mcitedefaultendpunct}{\mcitedefaultseppunct}\relax
\EndOfBibitem
\bibitem[Tan and García(2019)Tan, and García]{C7CS00813A}
Tan,~D.; García,~F. Main Group Mechanochemistry: from Curiosity to Established
  Protocols. \emph{Chem. Soc. Rev.} \textbf{2019}, \emph{48}, 2274--2292\relax
\mciteBstWouldAddEndPuncttrue
\mciteSetBstMidEndSepPunct{\mcitedefaultmidpunct}
{\mcitedefaultendpunct}{\mcitedefaultseppunct}\relax
\EndOfBibitem
\bibitem[Šepelák \latin{et~al.}(2012)Šepelák, Bégin-Colin, and
  Le~Caër]{C2DT30349C}
Šepelák,~V.; Bégin-Colin,~S.; Le~Caër,~G. Transformations in Oxides Induced
  by High-energy Ball-milling. \emph{Dalton Trans.} \textbf{2012}, \emph{41},
  11927--11948\relax
\mciteBstWouldAddEndPuncttrue
\mciteSetBstMidEndSepPunct{\mcitedefaultmidpunct}
{\mcitedefaultendpunct}{\mcitedefaultseppunct}\relax
\EndOfBibitem
\bibitem[Šepelák \latin{et~al.}(2013)Šepelák, Düvel, Wilkening, Becker,
  and Heitjans]{C2CS35462D}
Šepelák,~V.; Düvel,~A.; Wilkening,~M.; Becker,~K.-D.; Heitjans,~P.
  Mechanochemical Reactions and Syntheses of Oxides. \emph{Chem. Soc. Rev.}
  \textbf{2013}, \emph{42}, 7507--7520\relax
\mciteBstWouldAddEndPuncttrue
\mciteSetBstMidEndSepPunct{\mcitedefaultmidpunct}
{\mcitedefaultendpunct}{\mcitedefaultseppunct}\relax
\EndOfBibitem
\bibitem[Rak \latin{et~al.}(2014)Rak, Saadé, Friščić, and Moores]{Rak2014}
Rak,~M.~J.; Saadé,~N.~K.; Friščić,~T.; Moores,~A. Mechanosynthesis of
  Ultra-small Monodisperse Amine-stabilized Gold Nanoparticles with
  Controllable Size. \emph{Green Chem.} \textbf{2014}, \emph{16}, 86--89\relax
\mciteBstWouldAddEndPuncttrue
\mciteSetBstMidEndSepPunct{\mcitedefaultmidpunct}
{\mcitedefaultendpunct}{\mcitedefaultseppunct}\relax
\EndOfBibitem
\bibitem[Malca \latin{et~al.}(2017)Malca, Bao, Bastaille, Saad\'{e}, Kinsella,
  Fri{\v{s}}{\v{c}}i{\'c}, and Moores]{M2017bis}
Malca,~M.~Y.; Bao,~H.; Bastaille,~T.; Saad\'{e},~N.~K.; Kinsella,~J.~M.;
  Fri{\v{s}}{\v{c}}i{\'c},~T.; Moores,~A. Mechanically Activated Solvent-Free
  Assembly of Ultrasmall Bi2S3 Nanoparticles: a Novel, Simple, and Sustainable
  Means to Access Chalcogenide Nanoparticles. \emph{Chemistry of Materials}
  \textbf{2017}, \emph{29}, 7766--7773\relax
\mciteBstWouldAddEndPuncttrue
\mciteSetBstMidEndSepPunct{\mcitedefaultmidpunct}
{\mcitedefaultendpunct}{\mcitedefaultseppunct}\relax
\EndOfBibitem
\bibitem[Schreyer \latin{et~al.}(2019)Schreyer, Eckert, Immohr, de~Bellis,
  Felderhoff, and Sch{\"{u}}th]{https://doi.org/10.1002/anie.201903545}
Schreyer,~H.; Eckert,~R.; Immohr,~S.; de~Bellis,~J.; Felderhoff,~M.;
  Sch{\"{u}}th,~F. {Milling Down to Nanometers: A General Process for the
  Direct Dry Synthesis of Supported Metal Catalysts}. \emph{Angewandte Chemie
  International Edition} \textbf{2019}, \emph{58}, 11262--11265\relax
\mciteBstWouldAddEndPuncttrue
\mciteSetBstMidEndSepPunct{\mcitedefaultmidpunct}
{\mcitedefaultendpunct}{\mcitedefaultseppunct}\relax
\EndOfBibitem
\bibitem[Ralphs \latin{et~al.}(2015)Ralphs, Chansai, Hardacre, Burch, Taylor,
  and James]{Ralphs2015}
Ralphs,~K.; Chansai,~S.; Hardacre,~C.; Burch,~R.; Taylor,~S.~F.; James,~S.~L.
  {Mechanochemical Preparation of Ag Catalysts for the N-octane-SCR de-NOx
  Reaction: Structural and Reactivity Effects}. \emph{Catalysis Today}
  \textbf{2015}, \emph{246}, 198--206\relax
\mciteBstWouldAddEndPuncttrue
\mciteSetBstMidEndSepPunct{\mcitedefaultmidpunct}
{\mcitedefaultendpunct}{\mcitedefaultseppunct}\relax
\EndOfBibitem
\bibitem[Ralphs \latin{et~al.}(2014)Ralphs, D'Agostino, Burch, Chansai,
  Gladden, Hardacre, James, Mitchell, and Taylor]{Ralphs2014}
Ralphs,~K.; D'Agostino,~C.; Burch,~R.; Chansai,~S.; Gladden,~L.~F.;
  Hardacre,~C.; James,~S.~L.; Mitchell,~J.; Taylor,~S.~F. {Assessing the
  Surface Modifications Following the Mechanochemical Preparation of a Ag/Al2O3
  Selective Catalytic Reduction Catalyst}. \emph{Catalysis Science and
  Technology} \textbf{2014}, \emph{4}, 531--539\relax
\mciteBstWouldAddEndPuncttrue
\mciteSetBstMidEndSepPunct{\mcitedefaultmidpunct}
{\mcitedefaultendpunct}{\mcitedefaultseppunct}\relax
\EndOfBibitem
\bibitem[Amrute \latin{et~al.}(2021)Amrute, {De Bellis}, Felderhoff, and
  Sch{\"{u}}th]{Amrute2021}
Amrute,~A.~P.; {De Bellis},~J.; Felderhoff,~M.; Sch{\"{u}}th,~F.
  {Mechanochemical Synthesis of Catalytic Materials}. \emph{Chemistry – A
  European Journal} \textbf{2021}, \relax
\mciteBstWouldAddEndPunctfalse
\mciteSetBstMidEndSepPunct{\mcitedefaultmidpunct}
{}{\mcitedefaultseppunct}\relax
\EndOfBibitem
\bibitem[Zhu \latin{et~al.}(2017)Zhu, Yang, Gao, Zhang, Shi, Sun, Wang, and
  Zhang]{zhu2017solvent}
Zhu,~Z.-Y.; Yang,~Q.-Q.; Gao,~L.-F.; Zhang,~L.; Shi,~A.-Y.; Sun,~C.-L.;
  Wang,~Q.; Zhang,~H.-L. Solvent-free Mechanosynthesis of Composition-tunable
  Cesium Lead Halide Perovskite Quantum Dots. \emph{The journal of physical
  chemistry letters} \textbf{2017}, \emph{8}, 1610--1614\relax
\mciteBstWouldAddEndPuncttrue
\mciteSetBstMidEndSepPunct{\mcitedefaultmidpunct}
{\mcitedefaultendpunct}{\mcitedefaultseppunct}\relax
\EndOfBibitem
\bibitem[Boldyreva(2013)]{C3CS60052A}
Boldyreva,~E. Mechanochemistry of Inorganic and Organic Systems: What is
  Similar{,} What is Different? \emph{Chem. Soc. Rev.} \textbf{2013},
  \emph{42}, 7719--7738\relax
\mciteBstWouldAddEndPuncttrue
\mciteSetBstMidEndSepPunct{\mcitedefaultmidpunct}
{\mcitedefaultendpunct}{\mcitedefaultseppunct}\relax
\EndOfBibitem
\bibitem[Baig and Varma(2012)Baig, and Varma]{C1CS15204A}
Baig,~R. B.~N.; Varma,~R.~S. Alternative Energy Input: Mechanochemical{,}
  Microwave and Ultrasound-assisted Organic Synthesis. \emph{Chem. Soc. Rev.}
  \textbf{2012}, \emph{41}, 1559--1584\relax
\mciteBstWouldAddEndPuncttrue
\mciteSetBstMidEndSepPunct{\mcitedefaultmidpunct}
{\mcitedefaultendpunct}{\mcitedefaultseppunct}\relax
\EndOfBibitem
\bibitem[Do and Fri{\v{s}}{\v{c}}i{\'c}(2017)Do, and
  Fri{\v{s}}{\v{c}}i{\'c}]{doi:10.1021/acscentsci.6b00277}
Do,~J.-L.; Fri{\v{s}}{\v{c}}i{\'c},~T. Mechanochemistry: a Force of Synthesis.
  \emph{ACS Central Science} \textbf{2017}, \emph{3}, 13--19\relax
\mciteBstWouldAddEndPuncttrue
\mciteSetBstMidEndSepPunct{\mcitedefaultmidpunct}
{\mcitedefaultendpunct}{\mcitedefaultseppunct}\relax
\EndOfBibitem
\bibitem[Fiss \latin{et~al.}(2020)Fiss, Vu, Douglas, Do, Friščić, and
  Moores]{fiss2020solvent}
Fiss,~B.~G.; Vu,~N.-N.; Douglas,~G.; Do,~T.-O.; Friščić,~T.; Moores,~A.
  Solvent-Free Mechanochemical Synthesis of Ultrasmall Nickel Phosphide
  Nanoparticles and Their Application as a Catalyst for the Hydrogen Evolution
  Reaction (HER). \emph{ACS Sustainable Chemistry \& Engineering}
  \textbf{2020}, \emph{8}, 12014--12024\relax
\mciteBstWouldAddEndPuncttrue
\mciteSetBstMidEndSepPunct{\mcitedefaultmidpunct}
{\mcitedefaultendpunct}{\mcitedefaultseppunct}\relax
\EndOfBibitem
\bibitem[Gaffet \latin{et~al.}(1999)Gaffet, Bernard, Niepce, Charlot, Gras,
  Le~Ca{\"e}r, Guichard, Delcroix, Mocellin, and Tillement]{gaffet1999some}
Gaffet,~E.; Bernard,~F.; Niepce,~J.-C.; Charlot,~F.; Gras,~C.; Le~Ca{\"e}r,~G.;
  Guichard,~J.-L.; Delcroix,~P.; Mocellin,~A.; Tillement,~O. Some Recent
  Developments in Mechanical Activation and Mechanosynthesis. \emph{Journal of
  Materials Chemistry} \textbf{1999}, \emph{9}, 305--314\relax
\mciteBstWouldAddEndPuncttrue
\mciteSetBstMidEndSepPunct{\mcitedefaultmidpunct}
{\mcitedefaultendpunct}{\mcitedefaultseppunct}\relax
\EndOfBibitem
\bibitem[Silva \latin{et~al.}(2003)Silva, Pinheiro, Miranda, G{\'o}es, and
  Sombra]{silva2003structural}
Silva,~C.; Pinheiro,~A.; Miranda,~M.; G{\'o}es,~J.; Sombra,~A. Structural
  Properties of Hydroxyapatite Obtained by Mechanosynthesis. \emph{Solid State
  Sciences} \textbf{2003}, \emph{5}, 553--558\relax
\mciteBstWouldAddEndPuncttrue
\mciteSetBstMidEndSepPunct{\mcitedefaultmidpunct}
{\mcitedefaultendpunct}{\mcitedefaultseppunct}\relax
\EndOfBibitem
\bibitem[Moores(2018)]{MOORES201833}
Moores,~A. Bottom Up, Solid-phase Syntheses of Inorganic Nanomaterials by
  Mechanochemistry and Aging. \emph{Current Opinion in Green and Sustainable
  Chemistry} \textbf{2018}, \emph{12}, 33 -- 37\relax
\mciteBstWouldAddEndPuncttrue
\mciteSetBstMidEndSepPunct{\mcitedefaultmidpunct}
{\mcitedefaultendpunct}{\mcitedefaultseppunct}\relax
\EndOfBibitem
\bibitem[Di~Nardo \latin{et~al.}(2019)Di~Nardo, Hadad, Nguyen Van~Nhien, and
  Moores]{C9GC00304E}
Di~Nardo,~T.; Hadad,~C.; Nguyen Van~Nhien,~A.; Moores,~A. Synthesis of High
  Molecular Weight Chitosan from Chitin by Mechanochemistry and Aging.
  \emph{Green Chem.} \textbf{2019}, \emph{21}, 3276--3285\relax
\mciteBstWouldAddEndPuncttrue
\mciteSetBstMidEndSepPunct{\mcitedefaultmidpunct}
{\mcitedefaultendpunct}{\mcitedefaultseppunct}\relax
\EndOfBibitem
\bibitem[Baláž \latin{et~al.}(2013)Baláž, Achimovičová, Baláž, Billik,
  Cherkezova-Zheleva, Criado, Delogu, Dutková, Gaffet, Gotor, Kumar, Mitov,
  Rojac, Senna, Streletskii, and Wieczorek-Ciurowa]{C3CS35468G}
Baláž,~P.; Achimovičová,~M.; Baláž,~M.; Billik,~P.;
  Cherkezova-Zheleva,~Z.; Criado,~J.~M.; Delogu,~F.; Dutková,~E.; Gaffet,~E.;
  Gotor,~F.~J.; Kumar,~R.; Mitov,~I.; Rojac,~T.; Senna,~M.; Streletskii,~A.;
  Wieczorek-Ciurowa,~K. Hallmarks of Mechanochemistry: from Nanoparticles to
  Technology. \emph{Chem. Soc. Rev.} \textbf{2013}, \emph{42}, 7571--7637\relax
\mciteBstWouldAddEndPuncttrue
\mciteSetBstMidEndSepPunct{\mcitedefaultmidpunct}
{\mcitedefaultendpunct}{\mcitedefaultseppunct}\relax
\EndOfBibitem
\bibitem[Debnath \latin{et~al.}(2009)Debnath, Kim, and Geckeler]{B905260G}
Debnath,~D.; Kim,~S.~H.; Geckeler,~K.~E. The First Solid-phase Route to
  Fabricate and Size-tune Gold Nanoparticles at Room Temperature. \emph{J.
  Mater. Chem.} \textbf{2009}, \emph{19}, 8810--8816\relax
\mciteBstWouldAddEndPuncttrue
\mciteSetBstMidEndSepPunct{\mcitedefaultmidpunct}
{\mcitedefaultendpunct}{\mcitedefaultseppunct}\relax
\EndOfBibitem
\bibitem[de~Oliveira \latin{et~al.}(2020)de~Oliveira, Torresi, Emmerling, and
  Camargo]{D0TA05183G}
de~Oliveira,~P. F.~M.; Torresi,~R.~M.; Emmerling,~F.; Camargo,~P. H.~C.
  Challenges and Opportunities in the Bottom-up Mechanochemical Synthesis of
  Noble Metal Nanoparticles. \emph{J. Mater. Chem. A} \textbf{2020}, \emph{8},
  16114--16141\relax
\mciteBstWouldAddEndPuncttrue
\mciteSetBstMidEndSepPunct{\mcitedefaultmidpunct}
{\mcitedefaultendpunct}{\mcitedefaultseppunct}\relax
\EndOfBibitem
\bibitem[Rak \latin{et~al.}(2016)Rak, Fri{\v{s}}{\v{c}}i{\'c}, and
  Moores]{C6RA03711A}
Rak,~M.~J.; Fri{\v{s}}{\v{c}}i{\'c},~T.; Moores,~A. One-step{,} Solvent-free
  Mechanosynthesis of Silver Nanoparticle-infused Lignin Composites for Use as
  Highly Active Multidrug Resistant Antibacterial Filters. \emph{RSC Adv.}
  \textbf{2016}, \emph{6}, 58365--58370\relax
\mciteBstWouldAddEndPuncttrue
\mciteSetBstMidEndSepPunct{\mcitedefaultmidpunct}
{\mcitedefaultendpunct}{\mcitedefaultseppunct}\relax
\EndOfBibitem
\bibitem[{\v{S}}epel{\'{a}}k \latin{et~al.}(2007){\v{S}}epel{\'{a}}k, Heitjans,
  and Becker]{Sepelak2007}
{\v{S}}epel{\'{a}}k,~V.; Heitjans,~P.; Becker,~K.~D. {Nanoscale Spinel Ferrites
  Prepared by Mechanochemical Route : Thermal Stability and Size Dependent
  Magnetic Properties}. \emph{Journal of Thermal Analysis and Calorimetry}
  \textbf{2007}, \emph{90}, 93--97\relax
\mciteBstWouldAddEndPuncttrue
\mciteSetBstMidEndSepPunct{\mcitedefaultmidpunct}
{\mcitedefaultendpunct}{\mcitedefaultseppunct}\relax
\EndOfBibitem
\bibitem[Bal{\'{a}}{\v{z}} \latin{et~al.}(2003)Bal{\'{a}}{\v{z}},
  Boldi{\v{z}}{\'{a}}rov{\'{a}}, Godo{\v{c}}{\'{i}}kov{\'{a}}, and
  Brian{\v{c}}in]{Balaz2003}
Bal{\'{a}}{\v{z}},~P.; Boldi{\v{z}}{\'{a}}rov{\'{a}},~E.;
  Godo{\v{c}}{\'{i}}kov{\'{a}},~E.; Brian{\v{c}}in,~J. {Mechanochemical Route
  for Sulphide Nanoparticles Preparation}. \emph{Materials Letters}
  \textbf{2003}, \emph{57}, 1585--1589\relax
\mciteBstWouldAddEndPuncttrue
\mciteSetBstMidEndSepPunct{\mcitedefaultmidpunct}
{\mcitedefaultendpunct}{\mcitedefaultseppunct}\relax
\EndOfBibitem
\bibitem[Bal{\'{a}}{\v{z}} \latin{et~al.}(2014)Bal{\'{a}}{\v{z}},
  Bal{\'{a}}{\v{z}}, {\v{C}}aplovi{\v{c}}ov{\'{a}}, Zorkovsk{\'{a}},
  {\v{C}}aplovi{\v{c}}, and Psotka]{C3FD00117B}
Bal{\'{a}}{\v{z}},~P.; Bal{\'{a}}{\v{z}},~M.;
  {\v{C}}aplovi{\v{c}}ov{\'{a}},~M.; Zorkovsk{\'{a}},~A.;
  {\v{C}}aplovi{\v{c}},~L.; Psotka,~M. {The Dual Role of Sulfur-containing
  Amino Acids in the Synthesis of IV–VI Semiconductor Nanocrystals: a
  Mechanochemical Approach}. \emph{Faraday Discuss.} \textbf{2014}, \emph{170},
  169--179\relax
\mciteBstWouldAddEndPuncttrue
\mciteSetBstMidEndSepPunct{\mcitedefaultmidpunct}
{\mcitedefaultendpunct}{\mcitedefaultseppunct}\relax
\EndOfBibitem
\bibitem[de~Oliveira \latin{et~al.}(2020)de~Oliveira, Michalchuk, Buzanich,
  Bienert, Torresi, Camargo, and Emmerling]{D0CC03862H}
de~Oliveira,~P. F.~M.; Michalchuk,~A. A.~L.; Buzanich,~A.~G.; Bienert,~R.;
  Torresi,~R.~M.; Camargo,~P. H.~C.; Emmerling,~F. Tandem X-ray Absorption
  Spectroscopy and Scattering for in situ Time-resolved Monitoring of Gold
  Nanoparticle Mechanosynthesis. \emph{Chem. Commun.} \textbf{2020}, \emph{56},
  10329--10332\relax
\mciteBstWouldAddEndPuncttrue
\mciteSetBstMidEndSepPunct{\mcitedefaultmidpunct}
{\mcitedefaultendpunct}{\mcitedefaultseppunct}\relax
\EndOfBibitem
\bibitem[de~Oliveira \latin{et~al.}(2020)de~Oliveira, Michalchuk, Marquardt,
  Feiler, Prinz, Torresi, Camargo, and Emmerling]{D0CE00826E}
de~Oliveira,~P. F.~M.; Michalchuk,~A. A.~L.; Marquardt,~J.; Feiler,~T.;
  Prinz,~C.; Torresi,~R.~M.; Camargo,~P. H.~C.; Emmerling,~F. Investigating the
  Role of Reducing Agents on Mechanosynthesis of Au Nanoparticles.
  \emph{CrystEngComm} \textbf{2020}, \emph{22}, 6261--6267\relax
\mciteBstWouldAddEndPuncttrue
\mciteSetBstMidEndSepPunct{\mcitedefaultmidpunct}
{\mcitedefaultendpunct}{\mcitedefaultseppunct}\relax
\EndOfBibitem
\bibitem[Ma \latin{et~al.}(2014)Ma, Yuan, Bell, and James]{Ma2014}
Ma,~X.; Yuan,~W.; Bell,~S.~E.; James,~S.~L. {Better Understanding of
  Mechanochemical Reactions: Raman Monitoring Reveals Surprisingly Simple
  `Pseudo-fluid' Model for a Ball Milling Reaction}. \emph{Chemical
  Communications} \textbf{2014}, \emph{50}, 1585--1587\relax
\mciteBstWouldAddEndPuncttrue
\mciteSetBstMidEndSepPunct{\mcitedefaultmidpunct}
{\mcitedefaultendpunct}{\mcitedefaultseppunct}\relax
\EndOfBibitem
\bibitem[Katsenis \latin{et~al.}(2015)Katsenis, Pu{\v{s}}kari{\'{c}},
  {\v{S}}trukil, Mottillo, Julien, U{\v{z}}arevi{\'{c}}, Pham, Do, Kimber,
  Lazi{\'{c}}, Magdysyuk, Dinnebier, Halasz, and
  Fri{\v{s}}{\v{c}}i{\'{c}}]{Katsenis2015}
Katsenis,~A.~D.; Pu{\v{s}}kari{\'{c}},~A.; {\v{S}}trukil,~V.; Mottillo,~C.;
  Julien,~P.~A.; U{\v{z}}arevi{\'{c}},~K.; Pham,~M.~H.; Do,~T.~O.;
  Kimber,~S.~A.; Lazi{\'{c}},~P.; Magdysyuk,~O.; Dinnebier,~R.~E.; Halasz,~I.;
  Fri{\v{s}}{\v{c}}i{\'{c}},~T. {In situ X-ray Diffraction Monitoring of a
  Mechanochemical Reaction Reveals a Unique Topology Metal-organic Framework}.
  \emph{Nature Communications} \textbf{2015}, \emph{6}\relax
\mciteBstWouldAddEndPuncttrue
\mciteSetBstMidEndSepPunct{\mcitedefaultmidpunct}
{\mcitedefaultendpunct}{\mcitedefaultseppunct}\relax
\EndOfBibitem
\bibitem[Batzdorf \latin{et~al.}(2015)Batzdorf, Fischer, Wilke, Wenzel, and
  Emmerling]{Batzdorf2015}
Batzdorf,~L.; Fischer,~F.; Wilke,~M.; Wenzel,~K.~J.; Emmerling,~F. {Direct in
  situ Investigation of Milling Reactions Using Combined X-ray Diffraction and
  Raman Spectroscopy}. \emph{Angewandte Chemie - International Edition}
  \textbf{2015}, \emph{54}, 1799--1802\relax
\mciteBstWouldAddEndPuncttrue
\mciteSetBstMidEndSepPunct{\mcitedefaultmidpunct}
{\mcitedefaultendpunct}{\mcitedefaultseppunct}\relax
\EndOfBibitem
\bibitem[Kulla \latin{et~al.}(2017)Kulla, Fischer, Benemann, Rademann, and
  Emmerling]{Kulla2017}
Kulla,~H.; Fischer,~F.; Benemann,~S.; Rademann,~K.; Emmerling,~F. {The Effect
  of the Ball to Reactant Ratio on Mechanochemical Reaction Times Studied by:
  In situ PXRD}. \emph{CrystEngComm} \textbf{2017}, \emph{19}, 3902--3907\relax
\mciteBstWouldAddEndPuncttrue
\mciteSetBstMidEndSepPunct{\mcitedefaultmidpunct}
{\mcitedefaultendpunct}{\mcitedefaultseppunct}\relax
\EndOfBibitem
\bibitem[Akimbekov \latin{et~al.}(2017)Akimbekov, Katsenis, Nagabhushana,
  Ayoub, Arhangelskis, Morris, Fri{\v{s}}{\v{c}}i{\'{c}}, and
  Navrotsky]{Akimbekov2017}
Akimbekov,~Z.; Katsenis,~A.~D.; Nagabhushana,~G.~P.; Ayoub,~G.;
  Arhangelskis,~M.; Morris,~A.~J.; Fri{\v{s}}{\v{c}}i{\'{c}},~T.; Navrotsky,~A.
  {Experimental and Theoretical Evaluation of the Stability of True MOF
  Polymorphs Explains Their Mechanochemical Interconversions}. \emph{Journal of
  the American Chemical Society} \textbf{2017}, \emph{139}, 7952--7957\relax
\mciteBstWouldAddEndPuncttrue
\mciteSetBstMidEndSepPunct{\mcitedefaultmidpunct}
{\mcitedefaultendpunct}{\mcitedefaultseppunct}\relax
\EndOfBibitem
\bibitem[Kulla \latin{et~al.}(2018)Kulla, Haferkamp, Akhmetova, R{\"{o}}llig,
  Maierhofer, Rademann, and Emmerling]{Kulla2018}
Kulla,~H.; Haferkamp,~S.; Akhmetova,~I.; R{\"{o}}llig,~M.; Maierhofer,~C.;
  Rademann,~K.; Emmerling,~F. {In Situ Investigations of Mechanochemical
  One-Pot Syntheses}. \emph{Angewandte Chemie - International Edition}
  \textbf{2018}, \emph{57}, 5930--5933\relax
\mciteBstWouldAddEndPuncttrue
\mciteSetBstMidEndSepPunct{\mcitedefaultmidpunct}
{\mcitedefaultendpunct}{\mcitedefaultseppunct}\relax
\EndOfBibitem
\bibitem[Germann \latin{et~al.}(2020)Germann, Katsenis, Huski{\'{c}}, Julien,
  U{\v{z}}arevi{\'{c}}, Etter, Farha, Fri{\v{s}}{\v{c}}i{\'{c}}, and
  Dinnebier]{Germann2020}
Germann,~L.~S.; Katsenis,~A.~D.; Huski{\'{c}},~I.; Julien,~P.~A.;
  U{\v{z}}arevi{\'{c}},~K.; Etter,~M.; Farha,~O.~K.;
  Fri{\v{s}}{\v{c}}i{\'{c}},~T.; Dinnebier,~R.~E. {Real-Time in Situ Monitoring
  of Particle and Structure Evolution in the Mechanochemical Synthesis of
  UiO-66 Metal-Organic Frameworks}. \emph{Crystal Growth and Design}
  \textbf{2020}, \emph{20}, 49--54\relax
\mciteBstWouldAddEndPuncttrue
\mciteSetBstMidEndSepPunct{\mcitedefaultmidpunct}
{\mcitedefaultendpunct}{\mcitedefaultseppunct}\relax
\EndOfBibitem
\bibitem[Michalchuk \latin{et~al.}(2017)Michalchuk, Tumanov, Konar, Kimber,
  Pulham, and Boldyreva]{Michalchuk2017}
Michalchuk,~A.~A.; Tumanov,~I.~A.; Konar,~S.; Kimber,~S.~A.; Pulham,~C.~R.;
  Boldyreva,~E.~V. {Challenges of Mechanochemistry: Is In Situ Real-Time
  Quantitative Phase Analysis Always Reliable? A Case Study of Organic Salt
  Formation}. \emph{Advanced Science} \textbf{2017}, \emph{4}\relax
\mciteBstWouldAddEndPuncttrue
\mciteSetBstMidEndSepPunct{\mcitedefaultmidpunct}
{\mcitedefaultendpunct}{\mcitedefaultseppunct}\relax
\EndOfBibitem
\bibitem[Ferguson \latin{et~al.}(2019)Ferguson, Moyano, Tribello, Crawford,
  Bringa, James, Kohanoff, and {Del P{\'{o}}polo}]{Ferguson2019}
Ferguson,~M.; Moyano,~M.~S.; Tribello,~G.~A.; Crawford,~D.~E.; Bringa,~E.~M.;
  James,~S.~L.; Kohanoff,~J.; {Del P{\'{o}}polo},~M.~G. {Insights into
  Mechanochemical Reactions at the Molecular Level: Simulated Indentations of
  Aspirin and Meloxicam Crystals}. \emph{Chemical Science} \textbf{2019},
  \emph{10}, 2924--2929\relax
\mciteBstWouldAddEndPuncttrue
\mciteSetBstMidEndSepPunct{\mcitedefaultmidpunct}
{\mcitedefaultendpunct}{\mcitedefaultseppunct}\relax
\EndOfBibitem
\bibitem[McKissic \latin{et~al.}(2014)McKissic, Caruso, Blair, and
  Mack]{mckissic2014comparison}
McKissic,~K.~S.; Caruso,~J.~T.; Blair,~R.~G.; Mack,~J. Comparison of Shaking
  Versus Baking: Further Understanding the Energetics of a Mechanochemical
  Reaction. \emph{Green Chemistry} \textbf{2014}, \emph{16}, 1628--1632\relax
\mciteBstWouldAddEndPuncttrue
\mciteSetBstMidEndSepPunct{\mcitedefaultmidpunct}
{\mcitedefaultendpunct}{\mcitedefaultseppunct}\relax
\EndOfBibitem
\bibitem[Andersen and Mack(2017)Andersen, and Mack]{Andersen2017}
Andersen,~J.~M.; Mack,~J. {Decoupling the Arrhenius Equation: Via
  Mechanochemistry}. \emph{Chemical Science} \textbf{2017}, \emph{8},
  5447--5453\relax
\mciteBstWouldAddEndPuncttrue
\mciteSetBstMidEndSepPunct{\mcitedefaultmidpunct}
{\mcitedefaultendpunct}{\mcitedefaultseppunct}\relax
\EndOfBibitem
\bibitem[LaMer and Dinegar(1950)LaMer, and Dinegar]{doi:10.1021/ja01167a001}
LaMer,~V.~K.; Dinegar,~R.~H. Theory, Production and Mechanism of Formation of
  Monodispersed Hydrosols. \emph{Journal of the American Chemical Society}
  \textbf{1950}, \emph{72}, 4847--4854\relax
\mciteBstWouldAddEndPuncttrue
\mciteSetBstMidEndSepPunct{\mcitedefaultmidpunct}
{\mcitedefaultendpunct}{\mcitedefaultseppunct}\relax
\EndOfBibitem
\bibitem[Xia \latin{et~al.}(2017)Xia, Gilroy, Peng, and
  Xia]{doi:10.1002/anie.201604731}
Xia,~Y.; Gilroy,~K.~D.; Peng,~H.-C.; Xia,~X. Seed-Mediated Growth of Colloidal
  Metal Nanocrystals. \emph{Angewandte Chemie International Edition}
  \textbf{2017}, \emph{56}, 60--95\relax
\mciteBstWouldAddEndPuncttrue
\mciteSetBstMidEndSepPunct{\mcitedefaultmidpunct}
{\mcitedefaultendpunct}{\mcitedefaultseppunct}\relax
\EndOfBibitem
\bibitem[Colacino \latin{et~al.}(2018)Colacino, Carta, Pia, Porcheddu, Ricci,
  and Delogu]{Colacino2018}
Colacino,~E.; Carta,~M.; Pia,~G.; Porcheddu,~A.; Ricci,~P.~C.; Delogu,~F.
  {Processing and Investigation Methods in Mechanochemical Kinetics}. \emph{ACS
  Omega} \textbf{2018}, \emph{3}, 9196--9209\relax
\mciteBstWouldAddEndPuncttrue
\mciteSetBstMidEndSepPunct{\mcitedefaultmidpunct}
{\mcitedefaultendpunct}{\mcitedefaultseppunct}\relax
\EndOfBibitem
\bibitem[Urakaev and Boldyrev(2000)Urakaev, and Boldyrev]{urakaev2000mechanism}
Urakaev,~F.~K.; Boldyrev,~V. Mechanism and Kinetics of Mechanochemical
  Processes in Comminuting Devices: 1. Theory. \emph{Powder Technology}
  \textbf{2000}, \emph{107}, 93--107\relax
\mciteBstWouldAddEndPuncttrue
\mciteSetBstMidEndSepPunct{\mcitedefaultmidpunct}
{\mcitedefaultendpunct}{\mcitedefaultseppunct}\relax
\EndOfBibitem
\bibitem[Chen \latin{et~al.}(2010)Chen, Wang, Yang, Xu, Goh, Pan, and
  Chen]{chen2010measuring}
Chen,~G.; Wang,~Y.; Yang,~M.; Xu,~J.; Goh,~S.~J.; Pan,~M.; Chen,~H. Measuring
  Ensemble-averaged Surface-enhanced Raman Scattering in the Hotspots of
  Colloidal Nanoparticle Dimers and Trimers. \emph{Journal of the American
  Chemical Society} \textbf{2010}, \emph{132}, 3644--3645\relax
\mciteBstWouldAddEndPuncttrue
\mciteSetBstMidEndSepPunct{\mcitedefaultmidpunct}
{\mcitedefaultendpunct}{\mcitedefaultseppunct}\relax
\EndOfBibitem
\bibitem[Boldyrev and Avvakumov(1971)Boldyrev, and
  Avvakumov]{boldyrev1971mechanochemistry}
Boldyrev,~V.~V.; Avvakumov,~E.~G. Mechanochemistry of Inorganic Solids.
  \emph{Russian Chemical Reviews} \textbf{1971}, \emph{40}, 303--317\relax
\mciteBstWouldAddEndPuncttrue
\mciteSetBstMidEndSepPunct{\mcitedefaultmidpunct}
{\mcitedefaultendpunct}{\mcitedefaultseppunct}\relax
\EndOfBibitem
\bibitem[Gutman(1998)]{gutman1998mechanochemistry}
Gutman,~E.~M. \emph{Mechanochemistry of Materials}; Cambridge Int Science
  Publishing, 1998\relax
\mciteBstWouldAddEndPuncttrue
\mciteSetBstMidEndSepPunct{\mcitedefaultmidpunct}
{\mcitedefaultendpunct}{\mcitedefaultseppunct}\relax
\EndOfBibitem
\bibitem[Humphry-Baker \latin{et~al.}(2016)Humphry-Baker, Garroni, Delogu, and
  Schuh]{Humphry-Baker2016}
Humphry-Baker,~S.~A.; Garroni,~S.; Delogu,~F.; Schuh,~C.~A. {Melt-driven
  Mechanochemical Phase Transformations in Moderately Exothermic Powder
  Mixtures}. \emph{Nature Materials} \textbf{2016}, \emph{15}, 1280--1286\relax
\mciteBstWouldAddEndPuncttrue
\mciteSetBstMidEndSepPunct{\mcitedefaultmidpunct}
{\mcitedefaultendpunct}{\mcitedefaultseppunct}\relax
\EndOfBibitem
\bibitem[Butyagin(1971)]{butyagin1971kinetics}
Butyagin,~P.~Y. Kinetics and Nature of Mechanochemical Reactions. \emph{Russian
  Chemical Reviews} \textbf{1971}, \emph{40}, 901--915\relax
\mciteBstWouldAddEndPuncttrue
\mciteSetBstMidEndSepPunct{\mcitedefaultmidpunct}
{\mcitedefaultendpunct}{\mcitedefaultseppunct}\relax
\EndOfBibitem
\bibitem[Carta \latin{et~al.}(2020)Carta, Colacino, Delogu, and
  Porcheddu]{D0CP01658F}
Carta,~M.; Colacino,~E.; Delogu,~F.; Porcheddu,~A. Kinetics of Mechanochemical
  Transformations. \emph{Phys. Chem. Chem. Phys.} \textbf{2020}, \emph{22},
  14489--14502\relax
\mciteBstWouldAddEndPuncttrue
\mciteSetBstMidEndSepPunct{\mcitedefaultmidpunct}
{\mcitedefaultendpunct}{\mcitedefaultseppunct}\relax
\EndOfBibitem
\bibitem[Elder \latin{et~al.}(2007)Elder, Provatas, Berry, Stefanovic, and
  Grant]{PhysRevB.75.064107}
Elder,~K.~R.; Provatas,~N.; Berry,~J.; Stefanovic,~P.; Grant,~M. Phase-Field
  Crystal Modeling and Classical Density Functional Theory of Freezing.
  \emph{Phys. Rev. B} \textbf{2007}, \emph{75}, 064107\relax
\mciteBstWouldAddEndPuncttrue
\mciteSetBstMidEndSepPunct{\mcitedefaultmidpunct}
{\mcitedefaultendpunct}{\mcitedefaultseppunct}\relax
\EndOfBibitem
\bibitem[Ramakrishnan and Yussouff(1979)Ramakrishnan, and
  Yussouff]{PhysRevB.19.2775}
Ramakrishnan,~T.~V.; Yussouff,~M. First-Principles Order-parameter Theory of
  Freezing. \emph{Phys. Rev. B} \textbf{1979}, \emph{19}, 2775--2794\relax
\mciteBstWouldAddEndPuncttrue
\mciteSetBstMidEndSepPunct{\mcitedefaultmidpunct}
{\mcitedefaultendpunct}{\mcitedefaultseppunct}\relax
\EndOfBibitem
\bibitem[Greenwood \latin{et~al.}(2011)Greenwood, Ofori-Opoku, Rottler, and
  Provatas]{Greenwood2011}
Greenwood,~M.; Ofori-Opoku,~N.; Rottler,~J.; Provatas,~N. {Modeling Structural
  Transformations in Binary Alloys with Phase Field Crystals}. \emph{Physical
  Review B - Condensed Matter and Materials Physics} \textbf{2011}, \emph{84},
  1--10\relax
\mciteBstWouldAddEndPuncttrue
\mciteSetBstMidEndSepPunct{\mcitedefaultmidpunct}
{\mcitedefaultendpunct}{\mcitedefaultseppunct}\relax
\EndOfBibitem
\bibitem[Tegze \latin{et~al.}(2009)Tegze, Gr\'an\'asy, T\'oth, Podmaniczky,
  Jaatinen, Ala-Nissila, and Pusztai]{PhysRevLett.103.035702}
Tegze,~G.; Gr\'an\'asy,~L.; T\'oth,~G.~I.; Podmaniczky,~F.; Jaatinen,~A.;
  Ala-Nissila,~T.; Pusztai,~T. Diffusion-controlled Anisotropic Growth of
  Stable and Metastable Crystal Polymorphs in the Phase-Field Crystal Model.
  \emph{Phys. Rev. Lett.} \textbf{2009}, \emph{103}, 035702\relax
\mciteBstWouldAddEndPuncttrue
\mciteSetBstMidEndSepPunct{\mcitedefaultmidpunct}
{\mcitedefaultendpunct}{\mcitedefaultseppunct}\relax
\EndOfBibitem
\bibitem[van Teeffelen \latin{et~al.}(2009)van Teeffelen, Backofen, Voigt, and
  L{\"o}wen]{van2009derivation}
van Teeffelen,~S.; Backofen,~R.; Voigt,~A.; L{\"o}wen,~H. Derivation of the
  Phase-Field-Crystal Model for Colloidal Solidification. \emph{Physical Review
  E} \textbf{2009}, \emph{79}, 051404\relax
\mciteBstWouldAddEndPuncttrue
\mciteSetBstMidEndSepPunct{\mcitedefaultmidpunct}
{\mcitedefaultendpunct}{\mcitedefaultseppunct}\relax
\EndOfBibitem
\bibitem[Greenwood \latin{et~al.}(2010)Greenwood, Provatas, and
  Rottler]{PhysRevLett.105.045702}
Greenwood,~M.; Provatas,~N.; Rottler,~J. Free Energy Functionals for Efficient
  Phase Field Crystal Modeling of Structural Phase Transformations. \emph{Phys.
  Rev. Lett.} \textbf{2010}, \emph{105}, 045702\relax
\mciteBstWouldAddEndPuncttrue
\mciteSetBstMidEndSepPunct{\mcitedefaultmidpunct}
{\mcitedefaultendpunct}{\mcitedefaultseppunct}\relax
\EndOfBibitem
\bibitem[Greenwood \latin{et~al.}(2011)Greenwood, Rottler, and
  Provatas]{PhysRevE.83.031601}
Greenwood,~M.; Rottler,~J.; Provatas,~N. Phase-Field-Crystal Methodology for
  Modeling of Structural Transformations. \emph{Phys. Rev. E} \textbf{2011},
  \emph{83}, 031601\relax
\mciteBstWouldAddEndPuncttrue
\mciteSetBstMidEndSepPunct{\mcitedefaultmidpunct}
{\mcitedefaultendpunct}{\mcitedefaultseppunct}\relax
\EndOfBibitem
\bibitem[Fallah \latin{et~al.}(2012)Fallah, Stolle, Ofori-Opoku, Esmaeili, and
  Provatas]{fallah2012phase}
Fallah,~V.; Stolle,~J.; Ofori-Opoku,~N.; Esmaeili,~S.; Provatas,~N. Phase-Field
  Crystal Modeling of Early Stage Clustering and Precipitation in Metal Alloys.
  \emph{Physical Review B} \textbf{2012}, \emph{86}, 134112\relax
\mciteBstWouldAddEndPuncttrue
\mciteSetBstMidEndSepPunct{\mcitedefaultmidpunct}
{\mcitedefaultendpunct}{\mcitedefaultseppunct}\relax
\EndOfBibitem
\bibitem[Smith and Provatas(2017)Smith, and Provatas]{Smith2017}
Smith,~N.; Provatas,~N. {Generalization of the Binary Structural Phase Field
  Crystal Model}. \emph{Physical Review Materials} \textbf{2017}, \emph{1},
  1--10\relax
\mciteBstWouldAddEndPuncttrue
\mciteSetBstMidEndSepPunct{\mcitedefaultmidpunct}
{\mcitedefaultendpunct}{\mcitedefaultseppunct}\relax
\EndOfBibitem
\bibitem[Ofori-Opoku \latin{et~al.}(2012)Ofori-Opoku, Hoyt, and
  Provatas]{Ofori-Opoku2012}
Ofori-Opoku,~N.; Hoyt,~J.~J.; Provatas,~N. {Phase-Field-Crystal Model of Phase
  and Microstructural Stability in Driven Nanocrystalline Systems}.
  \emph{Physical Review E - Statistical, Nonlinear, and Soft Matter Physics}
  \textbf{2012}, \emph{86}, 1--6\relax
\mciteBstWouldAddEndPuncttrue
\mciteSetBstMidEndSepPunct{\mcitedefaultmidpunct}
{\mcitedefaultendpunct}{\mcitedefaultseppunct}\relax
\EndOfBibitem
\bibitem[Enrique and Bellon(2000)Enrique, and Bellon]{Enrique2000}
Enrique,~R.~A.; Bellon,~P. {Compositional Patterning in Systems Driven by
  Competing Dynamics of Different Length Scale}. \emph{Physical Review Letters}
  \textbf{2000}, \emph{84}, 2885--2888\relax
\mciteBstWouldAddEndPuncttrue
\mciteSetBstMidEndSepPunct{\mcitedefaultmidpunct}
{\mcitedefaultendpunct}{\mcitedefaultseppunct}\relax
\EndOfBibitem
\bibitem[{A. Enrique} and Bellon(1999){A. Enrique}, and Bellon]{A.Enrique1999}
{A. Enrique},~R.; Bellon,~P. {Phase Stability Under Irradiation in Alloys with
  a Positive Heat of Mixing: Effective Thermodynamics Description}.
  \emph{Physical Review B - Condensed Matter and Materials Physics}
  \textbf{1999}, \emph{60}, 14649--14659\relax
\mciteBstWouldAddEndPuncttrue
\mciteSetBstMidEndSepPunct{\mcitedefaultmidpunct}
{\mcitedefaultendpunct}{\mcitedefaultseppunct}\relax
\EndOfBibitem
\bibitem[Provatas and Elder(2010)Provatas, and Elder]{Provatas2010}
Provatas,~N.; Elder,~K. \emph{{Phase-Field Methods in Material Science and
  Engineering}}; 2010; pp 213--218\relax
\mciteBstWouldAddEndPuncttrue
\mciteSetBstMidEndSepPunct{\mcitedefaultmidpunct}
{\mcitedefaultendpunct}{\mcitedefaultseppunct}\relax
\EndOfBibitem
\bibitem[Elder \latin{et~al.}(2002)Elder, Katakowski, Haataja, and
  Grant]{PhysRevLett.88.245701}
Elder,~K.~R.; Katakowski,~M.; Haataja,~M.; Grant,~M. Modeling Elasticity in
  Crystal Growth. \emph{Phys. Rev. Lett.} \textbf{2002}, \emph{88},
  245701\relax
\mciteBstWouldAddEndPuncttrue
\mciteSetBstMidEndSepPunct{\mcitedefaultmidpunct}
{\mcitedefaultendpunct}{\mcitedefaultseppunct}\relax
\EndOfBibitem
\bibitem[Nakamuro \latin{et~al.}(2021)Nakamuro, Sakakibara, Nada, Harano, and
  Nakamura]{Nakamuro2021}
Nakamuro,~T.; Sakakibara,~M.; Nada,~H.; Harano,~K.; Nakamura,~E. {Capturing the
  Moment of Emergence of Crystal Nucleus from Disorder}. \emph{Journal of the
  American Chemical Society} \textbf{2021}, \emph{0}, 8--12\relax
\mciteBstWouldAddEndPuncttrue
\mciteSetBstMidEndSepPunct{\mcitedefaultmidpunct}
{\mcitedefaultendpunct}{\mcitedefaultseppunct}\relax
\EndOfBibitem
\end{mcitethebibliography}

\end{document}